\documentclass[a4paper,10pt,twoside]{cpc-hepnp}

\usepackage{multicol}
\usepackage{graphicx}
\usepackage{booktabs}
\usepackage{amssymb,bm,mathrsfs,bbm,amscd}
\usepackage[tbtags]{amsmath}
\usepackage{lastpage}

\newcommand{\nc}{\newcommand}       
\nc{\vc}[1] {\mbox{\boldmath $#1$}} 
\nc{\del}       {\partial}              
\nc{\bra}       {\langle}               
\nc{\ket}       {\rangle}               
\nc{\bras}[1]   {\langle #1|}           
\nc{\kets}[1]   {|#1\rangle}            
\nc{\mapleft}[1]{           
 \smash{\mathop{\,          %
  \hbox to 1.5cm{\rightarrowfill}\, }\limits_{#1}}}
\nc{\beq}     {\begin{eqnarray}}
\nc{\eeq}    {\end{eqnarray}}
\nc{\nn}      {\nonumber\\}
\nc{\vs}      {\vspace{-0.275cm}}
\nc{\fra}    {\frac{1}{2}}
\newcommand{\vev}[1]{\langle #1 \rangle}
\newcommand{\mcl}{\mathcal}
\newcommand{\f}[2]{\frac{d^#2 #1}{(2\pi)^#2}}

\begin{document}

\fancyhead[l]{Submitted to ``Chinese Physics C''}

\footnotetext[0]{Received 1 August 2014}

\title{Study of Hadrons Using the Gaussian Functional Method \\ in the O(4) Linear $\sigma$ Model\thanks{Supported by National Natural Science
Foundation of China (No.\,11205011 and No.\,11005007)}}

\author{%
      Hua-Xing Chen$^{1;1)}$\email{hxchen@buaa.edu.cn}%
\quad Shotaro Imai$^{2,3;2)}$\email{imai@nucl.sci.hokudai.ac.jp}%
\quad Hiroshi Toki$^{4;3)}$\email{toki@rcnp.osaka-u.ac.jp}%
\quad Li-Sheng Geng$^{1;4)}$\email{lisheng.geng@buaa.edu.cn}%
}
\maketitle

\address{%
$^1$ School of Physics and Nuclear Energy Engineering and International Research Center for Nuclei and Particles in the Cosmos, Beihang University, Beijing 100191, China
\\
$^2$ Department of Physics, Kyoto University, Kitashirakawa-oiwake, Sakyo, Kyoto 606-8502, Japan
\\
$^3$ Meme Media Laboratory, Hokkaido University, Sapporo, Hokkaido 060-8628, Japan
\\
$^4$ Research Center for Nuclear Physics, Osaka University, Ibaraki, Osaka 567-0047, Japan
\\
}

\begin{abstract}
We study properties of hadrons in the O(4) linear $\sigma$ model, where we take into account fluctuations of mesons around their mean field values using the Gaussian functional (GF) method. In the GF method we calculate dressed $\sigma$ and $\pi$ masses, where we include the effect of fluctuations of mesons to find a better ground state wave function than the mean field approximation.  Then we solve the Bethe-Salpeter equations and calculate physical $\sigma$ and $\pi$ masses.  We recover the Nambu-Goldstone theorem for the physical pion mass to be zero in the chiral limit.  The $\sigma$ meson is a strongly correlated meson-meson state, and seems to have a two meson composite structure.  We calculate $\sigma$ and $\pi$ masses as functions of temperature for both the chiral limit and explicit chiral symmetry breaking case.  We get similar behaviors for the physical $\sigma$ and $\pi$ masses as the case of the mean field approximation, but the coupling constants are much larger than the values of the case of the mean field approximation.
\end{abstract}



\footnotetext[0]{\hspace*{-3mm}\raisebox{0.3ex}{$\scriptstyle\copyright$}2013
Chinese Physical Society and the Institute of High Energy Physics
of the Chinese Academy of Sciences and the Institute
of Modern Physics of the Chinese Academy of Sciences and IOP Publishing Ltd}%

\begin{multicols}{2}

\section{Introduction}

The linear $\sigma$ model is used for the O(4) symmetry case to discuss chiral symmetry breaking and the appearance of Nambu-Goldstone bosons.  The mean field approximation (MFA) is often used for this discussion, but the MFA may not be justified for the case of large coupling constant.  An improvement of MFA is achieved by introducing fluctuations around the mean field values using the Gaussian functional (GF) method~\cite{barnes80}.  This method corresponds to the CJT method~\cite{cornwall74} and is sometimes called Hartree-Fock approximation for bosons.  The ground state energy is obtained by solving the Schr\"{o}dinger equation with the trial Gaussian wave functional in the GF method.
The mean field equation for the field vacuum expectation values and the mass-gap equations for the dressed meson masses are determined by the minimization condition of the energy, i.e., the one- and two-point Green function Schwinger-Dyson (SD) equations.

The resulting dressed mass of the Nambu-Goldstone boson is, however, not zero, even when the chiral symmetry is spontaneously broken in the chiral limit. This problem was studied by several authors in the GF framework~\cite{dmitrasinovic95, dmitrasinovic96, Dmitrasinovic:203yv, Nakamura:2012zzb, tsue99, tsue08}.  In order to recover the Nambu-Goldstone theorem, we have to solve the Bethe-Salpeter (BS) equations (or four-point Green function SD equations)~\cite{nakamura01, chen10}.  The physical meson masses are defined as the poles of the scattering matrices, while the dressed masses in the GF mass-gap equations are considered as single meson masses in the meson spectrum.  The single meson spectrum corresponds to the single particle spectrum of nucleons obtained in the Hartree-Fock (HF) approximation in nuclear physics.  We should then introduce the Random Phase approximation (RPA) after the HF procedure in order to recover the translational symmetry and the particle-hole excitation spectrum for nuclear many-body system.  The Nambu-Jona-Lasinio (NJL) model uses the same procedure for the mass generation of fermions and the appearance of the zero mass pseudo-scalar bosons~\cite{nambu61}.  This whole procedure was later studied by Okopi\'{n}ska using the optimized expansion (OE) method for the effective action (potential), where the second order derivatives of the classical fields provide the physical masses of the Nambu-Goldstone bosons~\cite{okopinska93, okopinska96}.  She was able to derive the BS equations in the OE method and showed that the Nambu-Goldstone theorem is recovered.  Nakamura and Dmitra\v sinovi\' c studied this problem in detail using the GF+BS methods for the O(4) case~\cite{nakamura01}. They did not, however, give numerical results for physical sigma and pion masses to compare with experimental values~\cite{Beringer2012}.  Tsue {\it et al.} formulated the BS equations using the linear response theory~\cite{tsue99}.  Lenaghan {\it et al.} studied the SU(3) case using the CJT method at finite temperature~\cite{lenaghan00}. However, they did not solve the BS equations for mesons. It remains to study that the Nambu-Goldstone boson masses vanish in the chiral limit at zero and finite temperatures and the scalar meson masses are reasonable in energy and their property, when the chiral symmetry is spontaneously broken both for O(4) and SU(3) cases.

There is another line of thought to overcome the problem of the Nambu-Goldstone theorem, following detailed studies of the linear $\sigma$ model in the CJT method by Nemoto {\it et al.}~\cite{nemoto00} and Mark\'{o} {\it et al.}~\cite{marko13}.  Knowing that the physical meson masses are to be obtained by the second order derivatives of the CJT effective action, the CJT effective action to be minimized is expressed in terms of the physical meson masses by carefully examining the renormalization procedure.  This method is further developed in the form of a symmetry-improved CJT (SI-CJT) action proposed recently by Pilaftsis and Teresi~\cite{pilaftsis13}.  In the SI-CJT method, instead of the minimization equation of the two-loop CJT action in terms of the mean field, a condition of making the Goldstone-boson mass being zero is imposed in the chiral limit. Using the SI-CJT method, they calculated the phase transition and the variation of the Higgs and Goldstone bosons at finite temperature for the O(2) case.  This SI-CJT method was then used for the O(4) case by Mao~\cite{mao14}.  Qualitatively similar results were obtained for the Higgs and Goldstone masses at finite temperature as those of the MFA and the order of the phase transition at finite temperature turned out to be second order.  The linear $\sigma$ model at large $N$ limit was studied using the CJT model by Petropoulos~\cite{petropoulos99}, where the phase transition was of second order, and the order of the phase transition was generally studied by Ogure and Sato~\cite{ogure99}.  Chiku and Hatsuda studied the meson properties at finite temperature in an optimized perturbation theory using order by order renormalization scheme~\cite{chiku98}.  In this case the order of the phase transition is first order opposing the case of the SI-CJT method.  Roh and Matsui studied the chiral phase transition at finite temperature in the one-loop level effective potential~\cite{roh08}.  The O(4) linear $\sigma$ model was studied by Tsue and Matsuda in the line similar to the present study~\cite{tsue08}.  There are renewed interests on this subject recently raised by the improved CJT method~\cite{marko13, mao14} and it is important to study further the GF+BS method on the order of phase transition and compare with the improved CJT approach.

In this paper we shall study hadron properties using the GF method for the O(4) linear $\sigma$ model, and then use the BS equations to demonstrate the Nambu-Goldstone theorem in the chiral limit~\cite{Schiff:1963zza, Rosen:1969dw, Symanzik:1981wd, Stevenson:1985zy, Stevenson:1986na, stevenson, brihaye, Kerman:1998vt}. To this end, we consider the O(4) linear $\sigma$ model as a low energy effective theory of QCD, and introduce the cut-off momentum to regularize the GF+BS equations.  We follow the recent study of the QCD Lagrangian using the Cho-Fadeev-Niemi (CFN) variables, where the high energy gluon mode acquires mass around 1$\sim$2~GeV and the NJL Lagrangian with confinement is obtained by integrating over the high energy mode~\cite{kondo11}.  The O(4) linear $\sigma$ model is then obtained by bosonization of the NJL Lagrangian using the auxiliary bosons for quark-antiquark composites with and without the confinement effect~\cite{eguchi76, imai13}.  We shall see the properties of mesons after solving the BS equations with a reasonable range of parameters in the linear $\sigma$ model Lagrangian.  We discuss the recovery of the chiral symmetry at finite temperature and the behavior of sigma and pion masses for both the chiral limit and explicit chiral symmetry breaking case.  Before studying the interesting but complicated flavor SU(3) case~\cite{chen10}, it is a reasonable step to study the flavor SU(2) case so that we know the amount of shifts of meson masses due to the BS equations and the interaction strengths in the linear $\sigma$ model Lagrangian.

This paper is organized as follows. In {Sec.~\ref{sec:gaussian}} we briefly introduce the O(4) linear $\sigma$ model and use the GF method to calculate the dressed masses of the $\sigma$ and $\pi$ mesons, denoted as $M_\sigma$ and $M_\pi$. These results are then used in {Sec.~\ref{sec:BS}} to calculate their physical masses using the Bethe-Salpeter equations, denoted as $m_\sigma$ and $m_\pi$. The single channel BS equation of the $\sigma$-$\pi$ scattering T-matrix gives the physical $\pi$ mass, and the coupled-channel BS equations of the $\sigma$-$\sigma$ and $\pi$-$\pi$ scattering T-matrix give the physical $\sigma$ mass. The numerical analyses are made for the both the chiral limit and explicit chiral symmetry breaking case, both at zero temperature in {Sec.~\ref{sec:numerical}}, and at finite temperature in {Sec.~\ref{sec:finite}}.  A summary of this work is presented in {Sec.~\ref{sec:summary}}.

\section{Gaussian functional method}
\label{sec:gaussian}

The Lagrangian density of the O(4) linear $\sigma$ model is
\begin{eqnarray}
{\cal L} &=&  \frac{1}{2} \left( \partial_{\mu} {\phi} \right)^2 - V(\phi^2) \, ,
\label{eq:lag}
\end{eqnarray}
where $\phi = (\phi_0,\phi_1,\phi_2,\phi_3) = (\sigma, \vec \pi)$ is a column vector and the potential $V(\phi^2)$ is
\begin{eqnarray}
V(\phi^2) &=& - {1 \over 2} \mu_0^2 \phi^2 + {\lambda_0 \over 4} \left( \phi^2 \right)^2 \, ,
\end{eqnarray}
which contains two parameters: the mass $\mu_0$ and the coupling constant $\lambda_0$.

The chiral symmetry can be both explicitly and spontaneously broken. The explicitly broken case can be expressed by adding the following term:
\begin{equation}
{\cal L}_{\chi SB} = - {\cal H}_{\chi SB} = \varepsilon \sigma \, ,
\end{equation}
This expression is suggested by the underlying Nambu-Jona-Lasinio (NJL) model, which is caused by the bare quark mass~\cite{nambu61, imai13}.  In the MFA, we introduce a fluctuation field  $s \equiv \sigma - \langle \sigma \rangle_{0 {\rm B}}$ and express the potential as,
\begin{eqnarray}
V (\sigma, \vec \pi) &=&
{1 \over 2} \left(m_{\sigma {\rm B}}^{2} s^{2} +
m_{\pi {\rm B}}^2 {\vec \pi}^{2}\right)
\nonumber\\  &&+
\lambda_0 v_B
s \left(s^2 + {\vec \pi}^{2}\right) +
{\lambda_0 \over 4}
\left(s^2 + {\vec \pi}^{2}\right)^{2} \, ,
\end{eqnarray}
where the vacuum expectation value $\langle \sigma \rangle_{0 {\rm B}}$ and masses of the $\sigma$ and $\pi$ mesons of the corresponding fluctuation fields are:
\begin{eqnarray}
\langle \sigma \rangle_{0 {\rm B}} &=& v_{\rm B} = f_{\pi {\rm B}}
= - {\varepsilon \over \mu_{0}^2}
+ \lambda_{0} {v_{\rm B}^{3} \over \mu_{0}^2}\,,
\label{eq:vev}\\
m_{\sigma {\rm B}}^2 &=& - \mu_{0}^2 + 3 \lambda_0 f_{\pi {\rm B}}^2\,,
\label{eq:msig} \\
m_{\pi {\rm B}}^2 &=& - \mu_{0}^2 + \lambda_0 f_{\pi {\rm B}}^2 =
{\varepsilon \over v_{\rm B}}\,.
\label{eq:mpi}
\end{eqnarray}
From Eq.\,\eqref{eq:mpi}, we find that the Nambu-Goldstone theorem is fulfilled in the chiral limit as the pion mass $m_{\pi {\rm B}}$ goes to zero for $\varepsilon\rightarrow 0$.

The MFA does not take into account radiative corrections and fluctuations around the mean fields, even for the case of large coupling constant.  A better and natural method is the Gaussian functional method to treat fluctuations around the mean fields~\cite{barnes80, nakamura01}.  We use the following Gaussian ground state wave functional:
\begin{eqnarray}
\nonumber \Psi_0[\phi] &=& \mcl{N}\exp\Big(-\frac{1}{4}\int d\bm{x} d\bm{y} [\phi_i(\bm{x})-\vev{\phi_i(\bm{x})}] \times
\\ && \times G^{-1}_{ij}(\bm{x},\bm{y})[\phi_j(\bm{y})-\vev{\phi_j(\bm{y})}]\Big) \, ,
\end{eqnarray}
where $\vev{\phi_i}$ is the vacuum expectation value of the $i$-th field ($i=0,\dots,3$), and we define the meson propagators as
\begin{eqnarray}
G_{ij}(\bm{x},\bm{y})=\frac{1}{2}\delta_{ij}\int \f{{\bm k}}{3}\frac{1}{\sqrt{\bm{k}^2+M_i^2}}e^{i\bm{k}\cdot(\bm{x}-\bm{y})} \, .
\end{eqnarray}
We use the gap equations for the dressed $\sigma$ and $\pi$ masses, denoted as $M_\sigma \equiv M_{i=0}$ and $M_\pi \equiv M_{i=1,2,3}$, respectively.  Here the ``dressed'' means that we still need to use the Bethe-Salpeter equations to calculate their physical masses, which will be done in the next section. The Hamiltonian is written as
\begin{eqnarray}
\mcl{H} &=& \int d\bm{y} \delta(\bm{y}-\bm{x})\Big(-\frac{1}{2}\frac{\delta^2}{\delta \phi_i(\bm{x})\phi_i(\bm{y})}
\nonumber\\  && +\frac{1}{2}\nabla_{x}\phi_i(\bm{x})\nabla_{y}\phi_i(\bm{y})+V(\phi^2)+\mcl{H}_{\chi SB}\Big) \, .
\end{eqnarray}

\end{multicols}
\ruleup

We write down the vacuum (ground-state) energy density:
\begin{eqnarray}
\label{eq:vee}
{\cal E}(M_{i}, \langle  \phi_{i} \rangle) &=&
 - \varepsilon \langle \phi_{0} \rangle - {1 \over 2} \mu_{0}^{2}
\langle {\bm \phi} \rangle^{2}
+ {\lambda_{0} \over 4}[\langle {\bm \phi} \rangle^{2}]^{2}
+ \sum_{i} [I_{1}(M_{i}) - {1 \over 2} \mu_{0}^{2}I_{0}(M_{i})
\nn&&- {1 \over 2} M_{i}^{2} I_{0}(M_{i})]
+
{\lambda_{0} \over 4} [
6 \sum_{i} \langle \phi_{i} \rangle^{2} I_{0}(M_{i})
+ 2 \sum_{i \neq j} \langle \phi_{i} \rangle^{2} I_{0}(M_{j})
\nn&&+
3 \sum_{i} I_{0}^{2}(M_{i})
+ 2 \sum_{i < j} I_{0}(M_{i}) I_{0}(M_{j})] \,,
\end{eqnarray}
where the two integrations $I_{0}(M_{i})$ and $I_{1}(M_{i})$ are:
\begin{eqnarray}
I_{0}(M_{i}) &=& {1 \over 2} \int {d^3 {\bf k} \over (2 \pi)^{3}}
{1 \over \sqrt{ {\bf k}^{2} + M_{i}^{2}}}
=
i \int {d^{4} k \over (2 \pi)^{4}}
{1 \over {k^{2} - M_{i}^{2} + i \epsilon}} \, ,
\label{eq:II0}
\\
I_{1}(M_{i}) &=& {1 \over 2} \int {d^3 {\bf k} \over (2
\pi)^{3}}
\sqrt{{\bf k}^{2} + M_{i}^{2}}
=
- {i \over 2} \int {d^{4} k \over (2\pi)^{4}}
\log \left(k^{2} - M_{i}^{2} + i \epsilon \right) +
{\rm const.}
\label{eq:II1}
\end{eqnarray}
\ruledown

\begin{multicols}{2}

\end{multicols}
\ruleup
\begin{center}
\includegraphics[width=12cm]{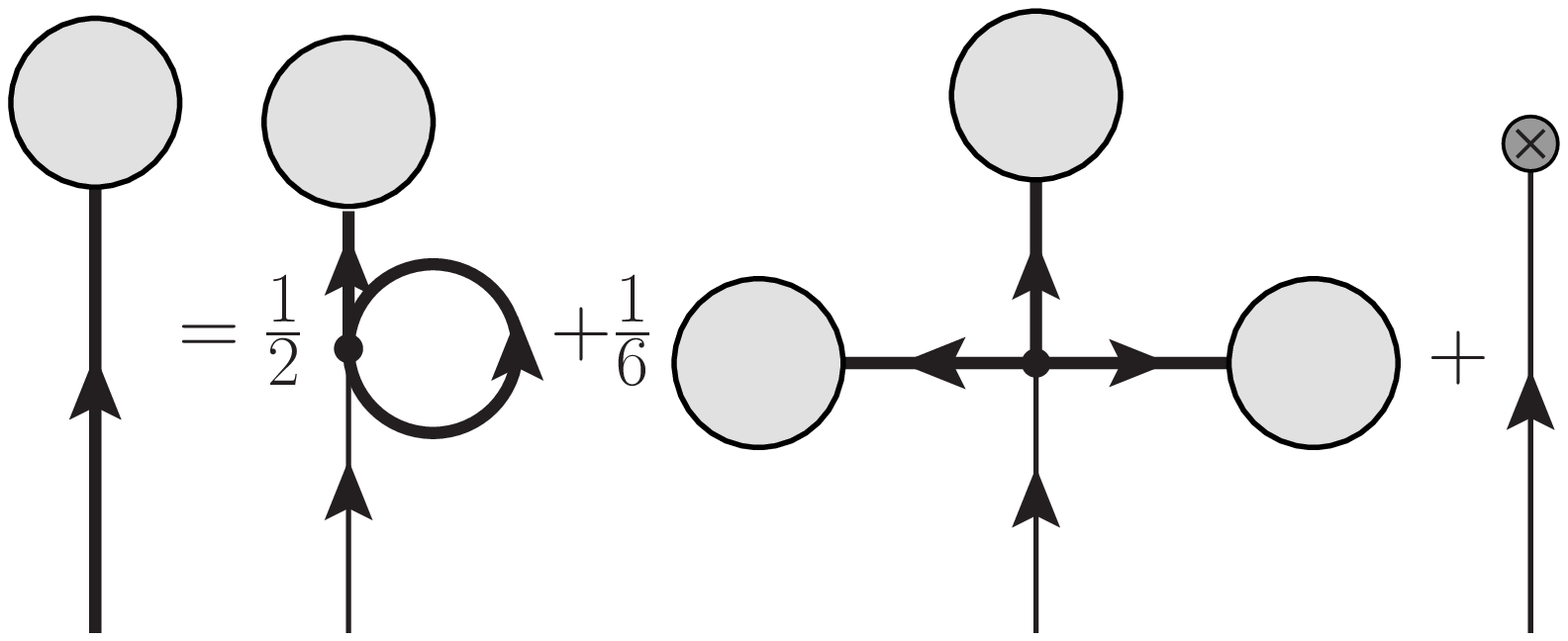}
\figcaption{\label{one} The diagrammatic expression of the one-point Green function Schwinger-Dyson equation. The thin solid line denotes the bare meson fields, the bold solid line is the dressed meson fields, the shaded blob together with the bold line is the vacuum expectation values of the field, and the solid dot in the interaction of the four lines denotes the bare four-point vertex. The diagrams are explicitly multiplied by their symmetry factors.}
\end{center}
\begin{center}
\includegraphics[width=12cm]{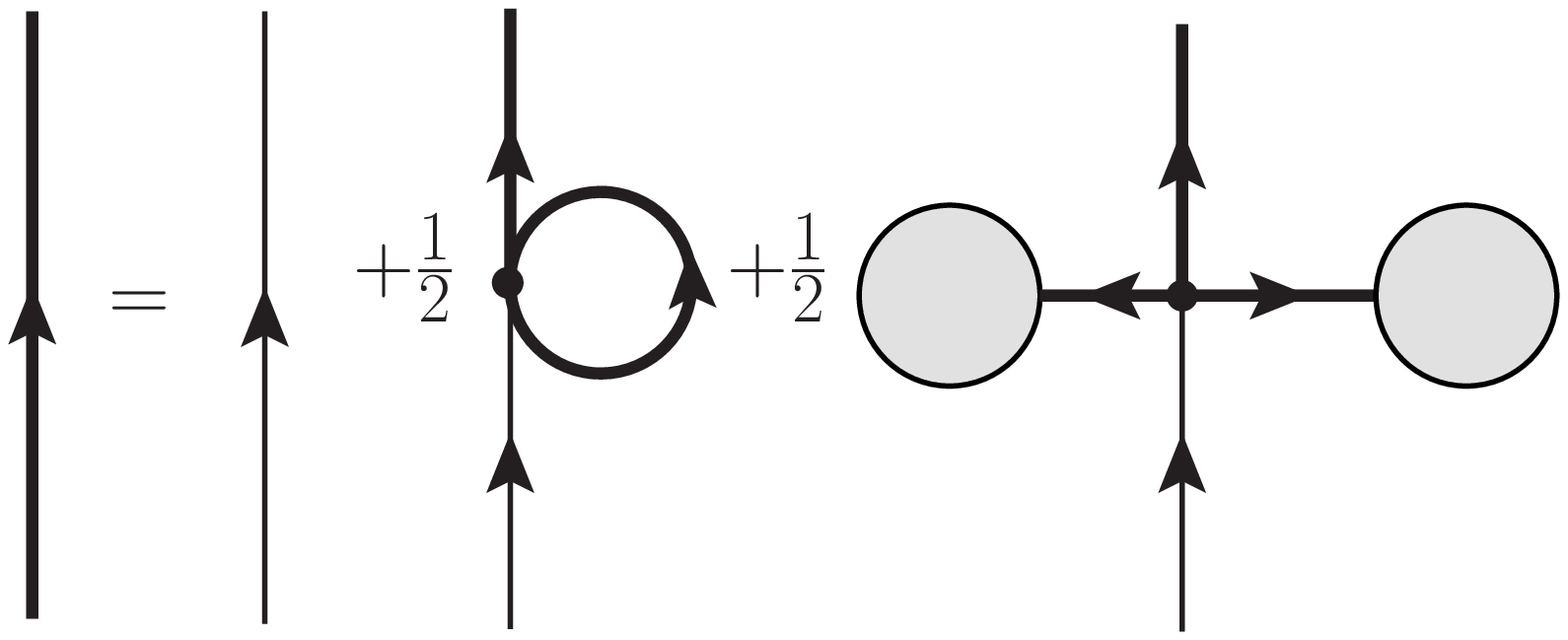}
\figcaption{\label{two} The diagrammatic expression of the two-point Green function Schwinger-Dyson equation with their symmetry factors.}
\end{center}
\ruledown

\begin{multicols}{2}

The field vacuum expectation value $\langle\phi_{i}\rangle$ and dressed $\pi$ and $\sigma$ masses can be obtained by the minimization condition, i.e., by varying the energy density~\eqref{eq:vee} with respect to the vacuum expectation values $\langle \phi_{i} \rangle$ and dressed meson masses $M_{i}$, respectively,
\begin{eqnarray}
\left(
{\partial {\cal E}(M_{i}, \langle  \phi_{i} \rangle)
\over
\partial \langle \phi_{i} \rangle,M_i}
\right)_{\text{min}} = 0 \, ,
\mbox{ for } i = 0 \ldots 3 .
\end{eqnarray}
These energy minimization conditions provide us with the following equations, which express the behaviors of spontaneous symmetry breaking in the Gaussian functional method:
\begin{eqnarray}
\langle \phi_{0} \rangle &=& v \,,
\label{eq:vevm}\\
\langle \phi_{i} \rangle &=& 0 ~  \mbox{ for } i = 1,2,3 \, ,
\label{eq:vevb}
\\ \mu_{0}^{2} &=& - {\varepsilon \over v} + {\lambda_{0}} \left[v^{2} + 3  I_{0}(M_\sigma) + 3  I_{0}(M_\pi) \right] \, ,
\label{eq:veva}
\\ M_\sigma^{2} &=& - \mu_{0}^{2} + {\lambda_{0}} \left[3 v^{2} + 3  I_{0}(M_\sigma) + 3  I_{0} (M_\pi) \right] \, ,
\label{eq:gapa}
\\ M_\pi^{2} &=& - \mu_{0}^{2} + {\lambda_{0}} \left[v^{2} +  I_{0}(M_\sigma) + 5  I_{0}(M_\pi) \right]
\label{eq:gapb} \, .
\end{eqnarray}
Note that Eq.\,\eqref{eq:veva} corresponds to the one-point Green function SD equation as shown in Fig.\,\ref{one}, and Eqs.\,\eqref{eq:gapa} and \eqref{eq:gapb} correspond to the two-point Green function SD equation as shown in Fig.\,\ref{two}.

The latter two equations can be further simplified by inserting the third equation \eqref{eq:veva}:
\begin{eqnarray}
M_\sigma^2 &=& {\varepsilon \over v} + 2 \lambda_0 v^{2} \, ,
\label{eq:gapM}
\\ M_\pi^2 &=& {\varepsilon \over v} + 2 \lambda_0  \left[ I_0(M_\pi) - I_0(M_\sigma) \right] \, .
\label{eq:gapm}
\end{eqnarray}
These two equations are convenient to obtain the dressed meson masses $M_{\sigma}$ and $M_{\pi}$, when we know the sigma mean field value $v \equiv \langle \phi_{0} \rangle$. From Eq.\,\eqref{eq:gapm}, we find that the Nambu-Goldstone theorem is not trivially fulfilled in the chiral limit any more as the pion mass $M_{\pi}$ does not simply (naturally) go to zero for $\varepsilon\rightarrow0$. However, if one solves the Bethe-Salpeter equations using the property of the polarization integral at zero energy (see Eq.\,\eqref{eq:regularization}) and calculates the physical pion mass, the Nambu-Goldstone theorem is recovered. This has been verified for both the $O(4)$ and $SU(3)$ cases~\cite{nakamura01,chen10}.
\section{Bethe-Salpeter equation}
\label{sec:BS}

In the previous section we have used the GF method to calculate the dressed $\sigma$ and $\pi$ masses, $M_\sigma$ and $M_\pi$.  To calculate their physical masses, $m_\sigma$ and $m_\pi$, we still need to solve BS equations, which correspond to find masses of the fluctuation fields. The single channel BS equation of the $\sigma$-$\pi$ scattering gives the physical pion mass $m_\pi$ and the coupled-channel BS equations of $\sigma$-$\sigma$ and $\pi$-$\pi$ scattering give the physical $\sigma$ mass $m_\sigma$. This method corresponds to the second order variation of the effective potential in terms of the fluctuation fields~\cite{okopinska96}. We note that by using the BS equation for the pseudoscalar fields the Nambu-Goldstone theorem can be fulfilled in the chiral limit due to the property of the polarization integral at zero energy~\eqref{eq:regularization}~\cite{nakamura01,chen10}.

\subsection{Single Channel $\sigma$-$\pi$ Scattering}
The single channel Bethe-Salpeter equation for the $\sigma$-$\pi$ scattering gives the physical $\pi$ mass. To do this, first we write the interaction kernel with the invariant energy $s=p^2$ in this channel:
\begin{eqnarray}
V_{\sigma\pi \rightarrow \sigma\pi}(s) &=& 2 \lambda_{0} \left[1 +
\left({2 \lambda_{0} v^{2} \over{s - M_\pi^{2}}}\right) \right] \, ,
\end{eqnarray}
which is shown diagrammatically in Fig.\,\ref{pot}.

With this interaction kernel we can get the T-matrix $T_{\sigma\pi\rightarrow\sigma\pi}(s)$ of the total four-point scattering amplitude $T(s,t,u)$ as
\begin{eqnarray}
 T_{\sigma\pi \rightarrow \sigma\pi}(s)
  = V_{\sigma\pi\rightarrow\sigma\pi}(s) + V_{\sigma\pi\rightarrow\sigma\pi}(s) G_{\sigma\pi\rightarrow\sigma\pi}(s) T_{\sigma\pi\rightarrow\sigma\pi}(s) \, ,
\label{eq:single}
\end{eqnarray}
which is shown diagrammatically in Fig.~\ref{bs}.

\end{multicols}
\ruleup
\begin{center}
\includegraphics[width=12cm]{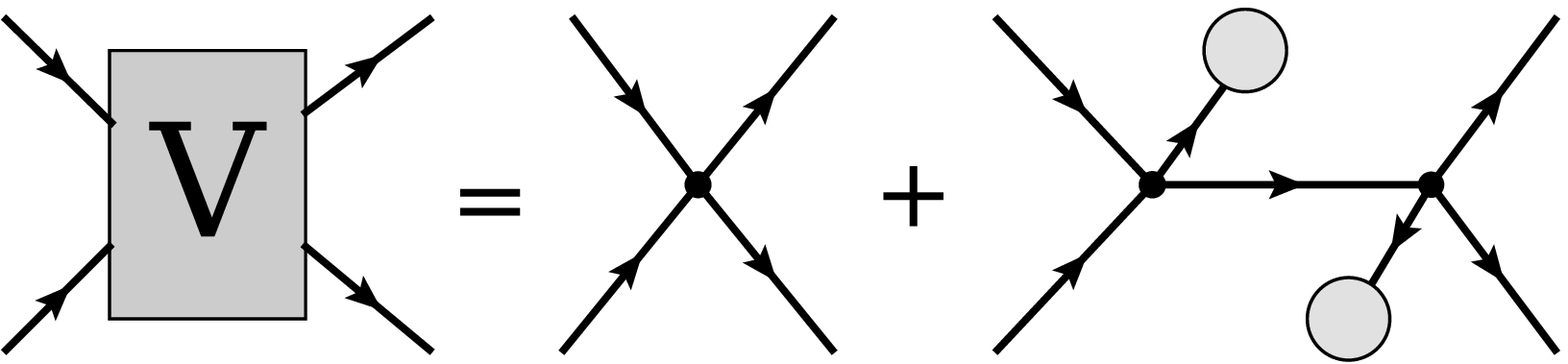}
\figcaption{\label{pot} The diagrammatic expression of the interaction kernel entering the Bethe-Salpeter equation.}
\end{center}
\begin{center}
\includegraphics[width=12cm]{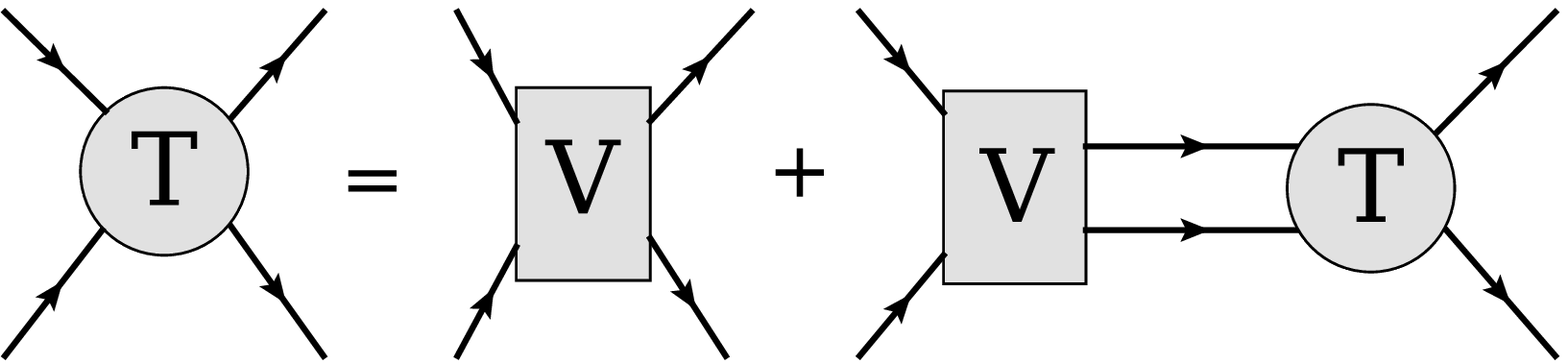}
\figcaption{\label{bs} The diagrammatic expression of the Bethe-Salpeter equation (or four-point Green function Schwinger-Dyson equation) with the T-matrix and the interaction kernel. All lines represent the dressed fields.}
\end{center}
\ruledown

\begin{multicols}{2}

The polarization function for meson masses is
\begin{eqnarray}
&&G_{\sigma\pi \rightarrow \sigma\pi}(s)
 \nn &&=
i  \int {d^{4} k \over (2 \pi)^{4}}
{1 \over {\left[k^{2} - M_\sigma^{2} + i \epsilon \right]
\left[(k - p)^{2} - M_\pi^{2} + i \epsilon \right]}} \, .
\label{eq:IMm}
\end{eqnarray}
The solution of Eq.~\eqref{eq:single} is simply expressed as
\begin{eqnarray}
T_{\sigma\pi\rightarrow\sigma\pi}(s) = {V_{\sigma\pi\rightarrow\sigma\pi}(s) \over{1 - V_{\sigma\pi\rightarrow\sigma\pi}(s) G_{\sigma\pi\rightarrow\sigma\pi}(s)}} \, .
\label{eq:BSpion}
\end{eqnarray}
The energy $s$, where the T-matrix diverges, corresponds to the pole condition:
\begin{eqnarray}
1 - V_{\sigma\pi\rightarrow\sigma\pi}(s) G_{\sigma\pi\rightarrow\sigma\pi}(s)=0
\end{eqnarray}
One can verify the Nambu-Goldstone theorem that the pseudoscalar meson has a zero mass in the chiral limit using this Bethe-Salpeter equation together with the mass gap equations~\eqref{eq:veva}-\eqref{eq:gapb} listed in {Sec.\,\ref{sec:gaussian}} and the property of the polarization integral at zero energy~\eqref{eq:regularization}. This has been done in Refs.~\cite{nakamura01,chen10}. In this paper we shall show this again numerically in {Sec.\,\ref{sec:numerical}}.

\subsection{Coupled Channel $\sigma$-$\sigma$ and $\pi$-$\pi$ Scattering}

The coupled channel Bethe-Salpeter equations of the $\sigma$-$\sigma$ and $\pi$-$\pi$ scattering give the physical $\sigma$ mass. To do this, first we write the interaction kernels in these channels:

\end{multicols}
\ruleup
\begin{eqnarray}
\bf{V} &=&
\left(\begin{array}{cc}
V_{\sigma \sigma\rightarrow \sigma \sigma} & V_{\sigma \sigma\rightarrow \pi \pi}
\\
V_{\pi \pi\rightarrow \sigma \sigma} & {1 \over 3}V_{\pi \pi\rightarrow \pi \pi}
\end{array}\right)
=
2 \lambda_{0}
\left(\begin{array}{cc}
3 \left[1 + 3 {2 \lambda_{0} v^{2} \over s - M_\sigma^{2}} \right] &
\left[1 + 3 {2 \lambda_{0} v^{2} \over s - M_\sigma^{2}} \right] \\
& \\
\left[1 + 3 {2 \lambda_{0} v^{2} \over s - M_\sigma^{2}} \right] &
{1 \over 3} \left[5 + 3 {2 \lambda_{0} v^{2} \over s - M_\sigma^{2}} \right] \\
\end{array}\right) \, ,
\end{eqnarray}
\ruledown

\begin{multicols}{2}

where we have used the matrix form. With this interaction kernel we get the T-matrices ${\bf T}(s)$ also in the matrix form:
\begin{eqnarray}
{\bf T} &=& {\bf V} + \frac{1}{2}{\bf V}{\bf G} {\bf T} \, ,
\label{eq:equationsigma}
\end{eqnarray}
where
\begin{eqnarray}
{\bf T} &=&
\left(\begin{array}{cc}
T_{\sigma \sigma\rightarrow \sigma \sigma} & T_{\sigma \sigma\rightarrow \pi \pi}
\\
T_{\pi \pi\rightarrow \sigma \sigma} & {1 \over 3}T_{\pi \pi\rightarrow \pi \pi}
\end{array}\right) ~ ,
\\
{\bf G} &=&
\left(\begin{array}{cc}
G_{\sigma\sigma\rightarrow \sigma\sigma} & 0 \\
0 & 3G_{\pi \pi\rightarrow \pi \pi}
\end{array}\right) \, .
\end{eqnarray}
The polarization functions $G_{\sigma\sigma\rightarrow\sigma\sigma}$ and $G_{\pi \pi\rightarrow\pi \pi}$ are given by
\begin{eqnarray}
&& G_{aa \rightarrow aa}(s)
\nn &=&
i  \int {d^{4} k \over (2 \pi)^{4}}
{1 \over {\left[k^{2} - M_a^2 + i \varepsilon
\right]
\left[(k - p)^{2} - M_a^2 + i \varepsilon
\right]}} \, .
\label{eq:Imm}
\end{eqnarray}
Here, $a$ denotes $\sigma$ and $\pi$.
The solution to the matrix equation \eqref{eq:equationsigma} is written in the matrix form:
\begin{eqnarray}
{\bf T} = (1 - \displaystyle \frac{1}{2}{\bf V}{\bf G})^{-1}{\bf V} \, ,
\label{eq:BSsigma}
\end{eqnarray}
As discussed in Ref.~\cite{nakamura01}, the discriminant of this equation can be simplified to be roots of
\begin{eqnarray}
(s - M_\sigma^2) \mathcal{T} (s) = 0 \, ,
\end{eqnarray}
where
\begin{eqnarray}
&& \mathcal{T} (s)
\\ \nonumber &=& 1 + 6 \lambda_0 G_{\sigma\sigma\rightarrow\sigma\sigma}(s) G_{\pi\pi\rightarrow\pi\pi}(s) V_{\sigma\sigma\rightarrow\pi\pi}(s)
\nn &-& {1\over2} \Big ( V_{\sigma\sigma\rightarrow\sigma\sigma}(s) G_{\sigma\sigma\rightarrow\sigma\sigma}(s) + V_{\pi\pi\rightarrow\pi\pi}(s) G_{\pi\pi\rightarrow\pi\pi}(s) \Big ) \, .
\end{eqnarray}

\section{Numerical results}
\label{sec:numerical}

In order to understand the GF+BS method, we first study the zero temperature case in this section, and the finite temperature case will be studied in the next section. Moreover, in this section we first study the chiral limit ($\varepsilon=0$) and then study the explicit chiral symmetry breaking case ($\varepsilon \neq 0$).

Here we give the general form of the necessary three-dimensional integrations, including the temperature $T=1/\beta$. We use the Matsubara formalism. The zero temperature integrals can be simply obtained by taking the limit $T \rightarrow 0$. First we write integrals for the Gaussian functional method in Eqs.\,\eqref{eq:II0} and \eqref{eq:II1}, including the temperature $T=1/\beta$:
\begin{eqnarray}
I_0(M_a,T)&=&-T\sum_{n}\int \f{k}{3}\frac{1}{(i\omega_n)^2-\omega_a^2}
\nn &=& \int \f{k}{3}\frac{1}{\omega_a}\left(\frac{1}{2}+\frac{1}{e^{\omega_a\beta}-1}\right) \, ,
\\ I_1(M_a,T)&=&-T\sum_n\int\f{k}{3}\ln((i\omega_n)^2-\omega_a^2)
\nn &=& \int\f{k}{3}\frac{1}{2}[\omega_a+T\ln(1-e^{-\omega_a\beta})] \, ,
\end{eqnarray}
where $\omega_a = \sqrt{{\bf k}^2 + M_a^2}$ and the Matsubara frequencies $\omega_n=2\pi nT$.

\end{multicols}
\ruleup

We write polarization integrals for the Bethe-Salpeter equations~\eqref{eq:IMm} and \eqref{eq:Imm}, also including the temperature $T=1/\beta$:
\begin{eqnarray}
G_{\sigma\pi \rightarrow \sigma\pi}(s,T)&=&\int\f{k}{3}\left[\frac{1}{2\omega_{\sigma}[(\omega_{\sigma}+\sqrt{s})^2-\omega_{\pi}^2]}
+\frac{1}{2\omega_{\pi}[(\omega_{\pi}-\sqrt{s})^2-\omega_{\sigma}^2]}\right.\nn
&&\left.+\frac{1}{2\omega_{\sigma}}\left(\frac{1}{(\omega_{\sigma}-\sqrt{s})^2-\omega_{\pi}^2}+\frac{1}{(\omega_{\sigma}+\sqrt{s})^2
-\omega_{\pi}^2}\right)\frac{1}{e^{\omega_{\sigma}\beta}-1}\right.\nn
&&\left.+\frac{1}{2\omega_{\pi}}\left(\frac{1}{(\omega_{\pi}+\sqrt{s})^2-\omega_{\sigma}^2}\frac{1}{e^{(\omega_{\pi}+\sqrt{s})\beta}-1}\right.\right.\nn&&\left.\left.
+\frac{1}{(\omega_{\pi}-\sqrt{s})^2-\omega_{\sigma}^2}\frac{1}{e^{(\omega_{\pi}-\sqrt{s})\beta}-1}\right)\right],\label{item}
\\
G_{aa \rightarrow aa}(s,T)&=&\int\f{k}{3}\left[\frac{1}{\omega_a(s-4\omega_a^2)}\left(1+\frac{1}{e^{\omega_a\beta}-1}\right)\right.\nn
&&\left.+\frac{1}{2\omega_a \sqrt{s}}\left(\frac{1}{\sqrt{s}+2\omega_a}\frac{1}{e^{(\omega_a+\sqrt{s})\beta}-1}+\frac{1}{\sqrt{s}-2\omega_a}\frac{1}{e^{(\omega_a-\sqrt{s})\beta}-1}\right)\right],
\end{eqnarray}
where $\omega_{\sigma}=\sqrt{\bm{k}^2+M_{\sigma}^2}$, $\omega_{\pi}=\sqrt{\bm{k}^2+M_{\pi}^2}$. They satisfy the following relations:
\begin{eqnarray}
G_{\sigma\pi\rightarrow\sigma\pi}(s,T) &\xrightarrow{M_\sigma \rightarrow M_a, M_\pi \rightarrow M_a}& G_{aa \rightarrow aa}(s,T) \, ,
\\ G_{\sigma\pi\rightarrow\sigma\pi}(s,T) &\xrightarrow{~~~~~~s \rightarrow 0~~~~~~}& {I_0(M_\sigma,T) - I_0(M_\pi,T) \over M_\sigma^2 - M_\pi^2} \, .
\label{eq:regularization}
\end{eqnarray}
\ruledown

\begin{multicols}{2}

We show explicitly the Nambu-Goldstone theorem in Eq. \eqref{eq:regularization} for the GFM case.  We note that this is an equation relating one-particle loop and two-particle loop. There may be similar equations relating these loops to multi-particle loops which can be important to recover the Nambu-Goldstone theorem when including the higher-order effects.
We note that Eq. \eqref{eq:regularization} is important to recover the Nambu-Goldstone theorem when solving Bethe-Salpeter equations~\cite{nakamura01,chen10}.  This means that the above relation of $G_{\sigma\pi\rightarrow\sigma\pi}(s,T)$ with $I_0(M_\sigma,T) - I_0(M_\pi,T)$ should not be broken in any regularization and renormalization procedures.

These loop integrals can be separated into two parts: one part contains the temperature $T$, that is finite, and the other does not contain it, that is infinite.  Therefore, we have to regularize the temperature independent part, which needs some discussion.   We consider that the linear $\sigma$ model is a low energy effective theory, which is obtained by the bosonization of the NJL Lagrangian using the auxiliary boson fields for quark-antiquark states~\cite{imai13, eguchi76}.  The NJL model is also a low energy effective theory, which is obtained by integrating out the high energy mode $\chi$ of the Cho-Niemi-Fadeev variables in the QCD Lagrangian, shown by Kondo~\cite{kondo11}.  Hence, we naturally have a physical scale on the order of the mass of $\chi$ (1 $\sim$ 2 GeV), and introduce here a three-dimensional cut-off momentum $\Lambda$ to represent the low energy scale.   For the temperature dependent part, we take the integral over momentum up to infinity.  Even if we introduce $\Lambda$ in the temperature dependent part, the results are changed only slightly and the present discussions are qualitatively unchanged.

\subsection{Chiral Limit ($\varepsilon=0$)}

First we consider the chiral limit, $\varepsilon=0$.  In Ref.~\cite{nakamura01} the authors have explicitly shown that the pion decay constant $f_\pi$ is similar to the sigma mean field value, $f_\pi \approx v$ (see their Eq.\,(4.15)) using the axial Ward-Takahashi identity, even for the correlated pions.
Therefore, we fix the sigma mean field value at $v=93$\,MeV for zero temperature at the beginning.

The linear $\sigma$ model Lagrangian contains three parameters; $\lambda_0$, $\mu_0$ and $\Lambda$ in the chiral limit.  The constraint that the mean field value is fixed to $v=93$\,MeV provides the relation among these three parameters.  We shall vary the coupling constant $\lambda_0$ to present numerical results.  We show first the relation between dressed $\sigma$ and $\pi$ masses, $M_\sigma$ and $M_\pi$, for various cut off $\Lambda$ in a large mass range in Fig.~\ref{fig:gapNoB}.

\begin{center}
\includegraphics[width=8cm]{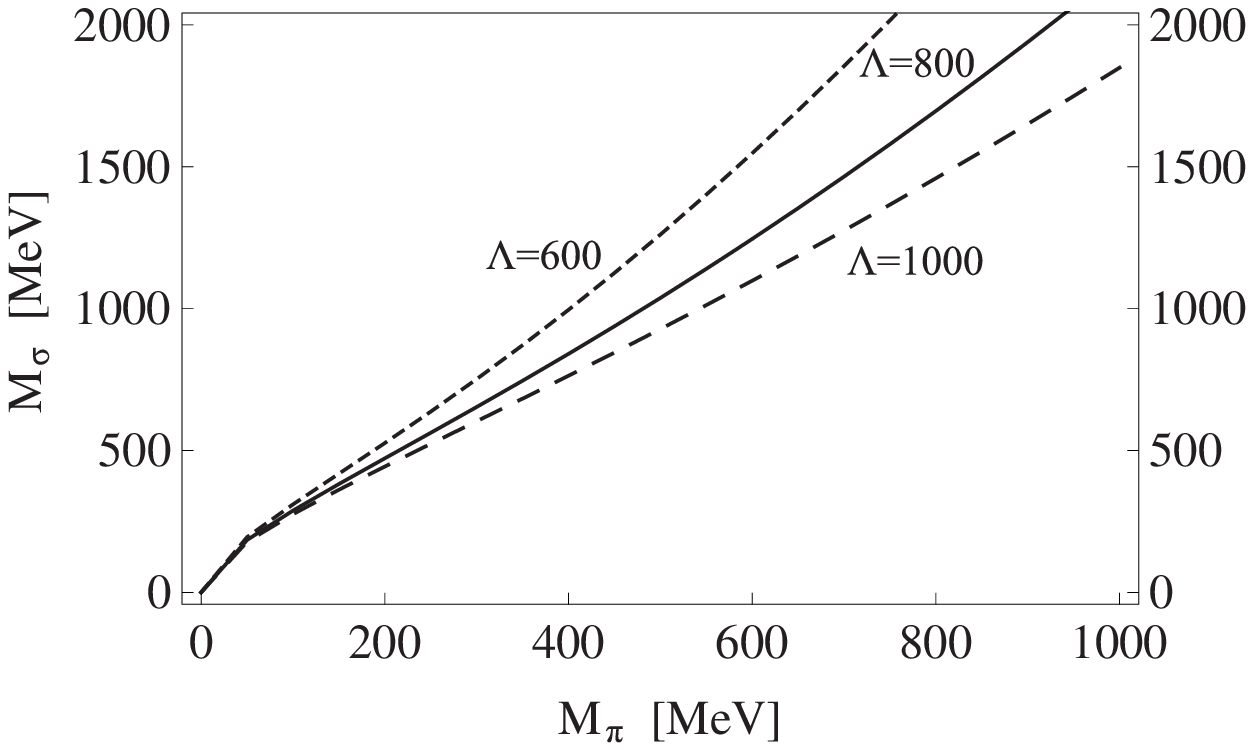}
\figcaption{\label{fig:gapNoB} The relation of the dressed $\sigma$ mass $M_\sigma$ and the dressed $\pi$ mass $M_\pi$, fixing the sigma mean field value at $v=93$\,MeV. The short-dashed, solid and long-dashed curves are for $\Lambda = 600$, $800$ and $1000$\,MeV, respectively. The calculations are done in the chiral limit.}
\end{center}

Generally $M_\sigma$ is larger than $M_\pi$, and the ratio between $M_\sigma$ and $M_\pi$ decreases as $\Lambda$ increases.  In order to understand the behaviors of these masses and their relations to the original Lagrangian, we plot $\mu_0$ as a function of $\lambda_0$ in the left panel of Fig.\,\ref{fig:NoB}, and the dressed masses of the $\sigma$ and $\pi$ mesons, $M_{\sigma}$ and $M_{\pi}$, also as functions of $\lambda_0$ in the right panel of Fig.\,\ref{fig:NoB}, fixing $\Lambda = 800$\,MeV. $M_{\sigma}$ and $M_{\pi}$ both increase with $\lambda_{0}$. We would like to point out that the dressed pion mass $M_{\pi}$ is nonzero, when the coupling constant $\lambda_{0}$ is finite.  This means that we need to solve Bethe-Salpeter equations in order to get the zero physical pion mass in the chiral limit.

\end{multicols}
\ruleup
\begin{center}
\includegraphics[width=8cm]{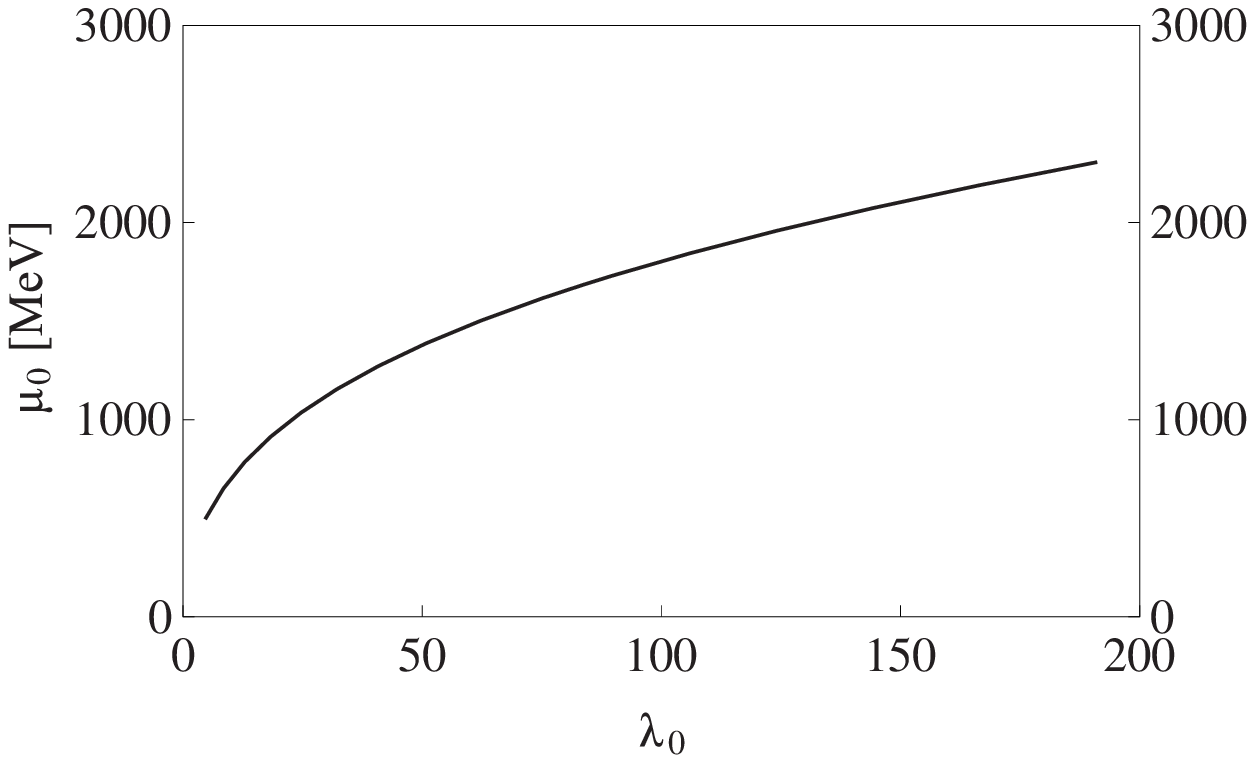}
\includegraphics[width=8cm]{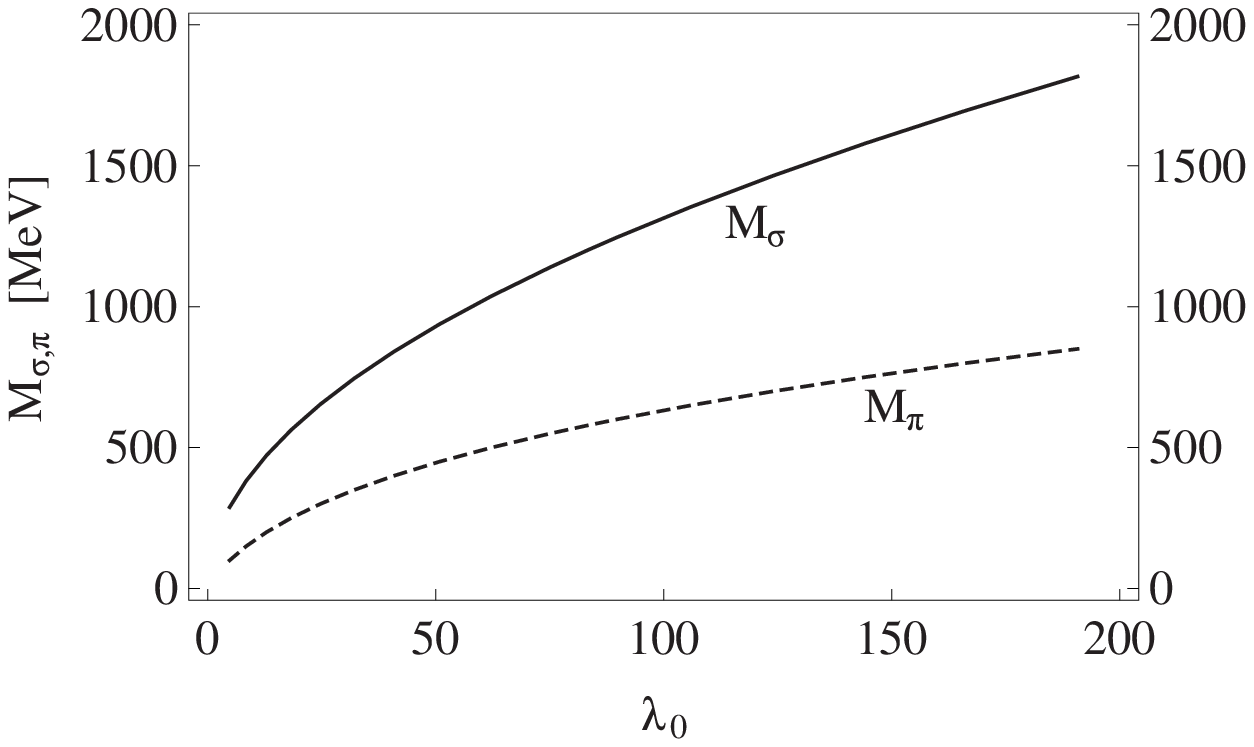}
\figcaption{\label{fig:NoB} In the left figure, we show $\mu_0$ as a function of $\lambda_0$. In the right figure, we show dressed $\sigma$ and $\pi$ masses, $M_\sigma$ and $M_\pi$, as functions of $\lambda_{0}$. We fix $\Lambda = 800$\,MeV. The sigma and pion masses go to zero as $\lambda_{0}\rightarrow 0$. The calculations are done in the chiral limit.}
\end{center}
\ruledown

\begin{multicols}{2}

We then plot the physical sigma and pion masses, $m_\sigma$ and $m_\pi$, as functions of the dressed pion mass $M_{\pi}$ in Fig.\,\ref{fig:BSNoB} by solving the Bethe-Salpeter equations \eqref{eq:BSpion} and \eqref{eq:BSsigma}. The physical pion mass is zero throughout, which means that the Nambu-Goldstone theorem is recovered after solving the Bethe-Salpeter equations. The physical sigma mass $m_\sigma$ increases with $M_{\pi}$. We find that the energy gain of the physical sigma and pion masses from their dressed masses are very large and of order $M_{\pi}$.

\begin{center}
\includegraphics[width=8cm]{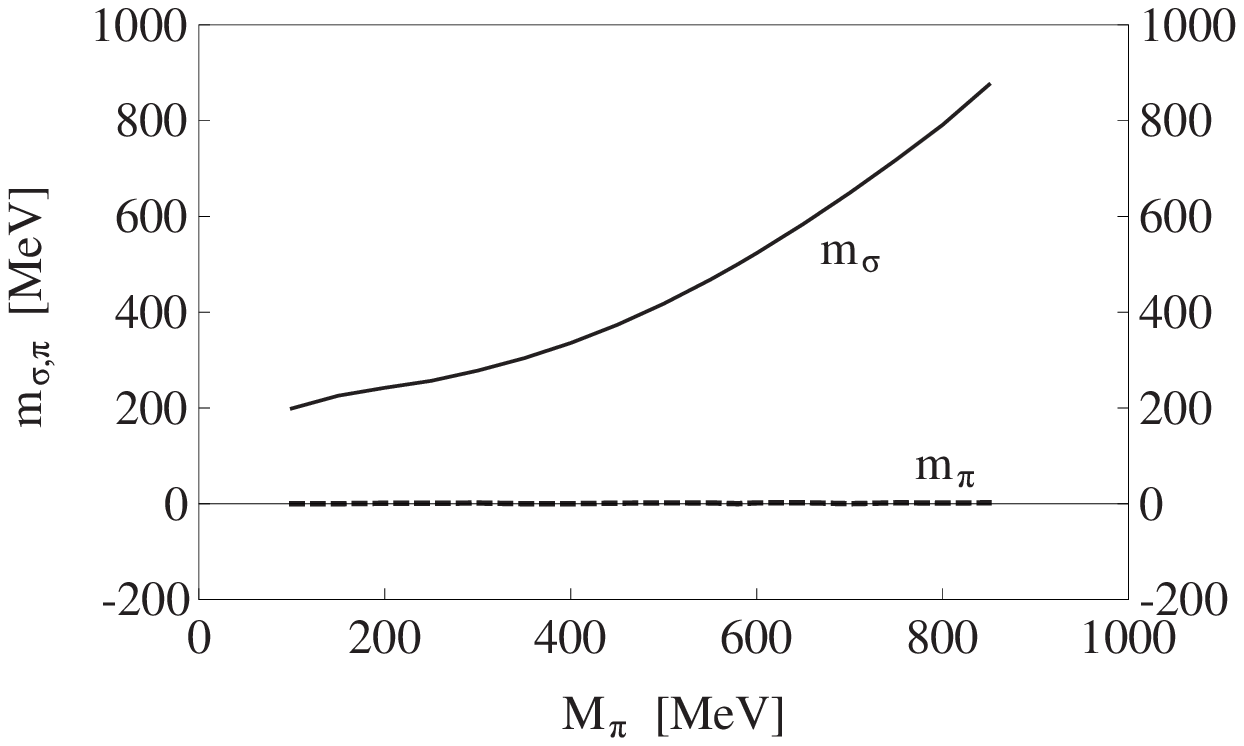}
\figcaption{\label{fig:BSNoB} Physical $\sigma$ and $\pi$ masses, $m_\sigma$ and $m_\pi$, as functions of $M_\pi$ for $\Lambda = 800$\,MeV using the Bethe-Salpeter equations (\ref{eq:BSpion}) and (\ref{eq:BSsigma}).  The physical pion mass $m_\pi$ stays zero when $M_{\pi}$ is finite. The calculations are done in the chiral limit.}
\end{center}

The physical sigma mass is around 500\,MeV~\cite{Beringer2012} and therefore we choose the parameters in the linear $\sigma$ model Lagrangian as:
\begin{eqnarray}
\lambda_0 &=& 83.6 \, ,
\nn
\mu_0 &=& 1680 ~{\rm MeV} \, ,
\nn
 \Lambda &=& 800 ~{\rm MeV} \, .\label{eq:NBparameter}
\end{eqnarray}
These values provide
\begin{eqnarray}
 v(T=0) &=& f_\pi = 93.0 \,{\rm MeV} \, ,
\nn
M_\sigma(T=0) &=& 1200 \,{\rm MeV} \, ,
\nn
M_\pi(T=0) &=& 580 \,{\rm MeV} \, ,
\\\label{eq:NBvalues}
m_\sigma(T=0) &=& 500 \,{\rm MeV} \, ,
\nn
m_\pi(T=0) &=& 0.0 \,{\rm MeV} \nonumber\, .
\end{eqnarray}
We find that in order to obtain the physical sigma mass $m_\sigma$ around its experimental value 500 MeV~\cite{Beringer2012}, we need to take a large coupling constant  $\lambda_0 = 83.6$.  It is interesting to point out that the pion and the sigma mesons both gain large energies of about 600$\sim$700\,MeV from their dressed masses due to the residual interaction treated by the Bethe-Salpeter equations.

We discuss here the nature of these physical $\sigma$ and $\pi$ mesons.  First of all the $\sigma$ model Lagrangian has a origin from quark dynamics as the Lagrangian can be derived from the NJL model by bosonization using mesons as auxiliary fields of quark-antiquark states~\cite{eguchi76, imai13}.  The GF method provides the optimum single meson (quark-antiquark state) spectra by variation as the Hartree-Fock method provides the optimum single particle states of nucleons for nucleus.  The BS method then provides physical meson states through interactions among meson-meson states and with single meson states.  The threshold energy of meson-meson states with spin-parity $0^+$ consisting of $\pi-\pi$ states is $M_\pi+M_\pi=$1160\,MeV, which is below the single sigma state at $M_\sigma=$1200\,MeV.  The interaction among $\pi-\pi$ states pushes down one state at 500\,MeV and the lowest scalar state should be a strongly correlated $\pi-\pi$ state, that indicates the lowest scalar state as a four quark state~\cite{chen10a}.  On the other hand, in the pseudoscalar channel the single pion state has the energy of $M_\pi=$580\,MeV, while $\sigma-\pi$ configuration starts at the energy of $M_\pi+M_\sigma=$1780\,MeV.  The interaction among these configurations brings down one state at zero energy and hence the physical pion should have a mixed nature of pion and pion-sigma states.  These features of compositeness should eventually lead to the non-linear $\sigma$ model expression of chiral dynamics.

It is interesting to point out that the BS equations do not develop imaginary terms for the $\sigma$ meson, since the physical masses are smaller than the dressed masses.  This is completely different from the mean field case, where the BS equation for the sigma meson provides imaginary term, which corresponds to the decay of the sigma meson into two pions.  It is our future work to calculate the transition strength of a composite sigma meson into two composite pions.  We should further study the properties of the physical pion in detail.

Based on these values, we shall study finite temperature effects in the case of chiral limit in {Sec.\,\ref{sec:finite}}, and we have indicated those observables at zero temperature with ``$T=0$'' in Eqs.\,\eqref{eq:NBvalues}, which depend on temperature.  Before going to finite temperature, we discuss the case of the explicit chiral symmetry breaking in the next subsection.

\subsection{Explicit chiral symmetry breaking case ($\varepsilon \neq 0$)}

The explicit chiral symmetry breaking is due to the term $\mathcal{L}_{\chi SB} = \varepsilon \sigma$, and we have an extra free parameter $\varepsilon$ in the model Lagrangian.  As suggested in Ref.~\cite{nakamura01}, we use the following expression
\begin{eqnarray}
\varepsilon = f_\pi m_{\pi0}^2 \, ,
\label{eq:epsilon}
\end{eqnarray}
where the parameter $m_{\pi0}$ is fine-tuned to be $m_{\pi0} = 142$\,MeV, so that the physical pion mass is calculated to be around its physical value $138$\,MeV~\cite{Beringer2012} by solving the Bethe-Salpeter equation.

To do the numerical analysis, we follow the same procedures as the case of chiral limit. Again we fix $v(\approx f_\pi)=93$\,MeV. The relation of $M_\sigma$ and $M_\pi$ is shown in Fig.\,\ref{fig:gapESB} for three cut-off momenta $\Lambda=600$, $800$ and $1000$\,MeV, which can be compared with the results of Ref.~\cite{nakamura01} as well as Fig.\,\ref{fig:gapNoB}.

\begin{center}
\includegraphics[width=8cm]{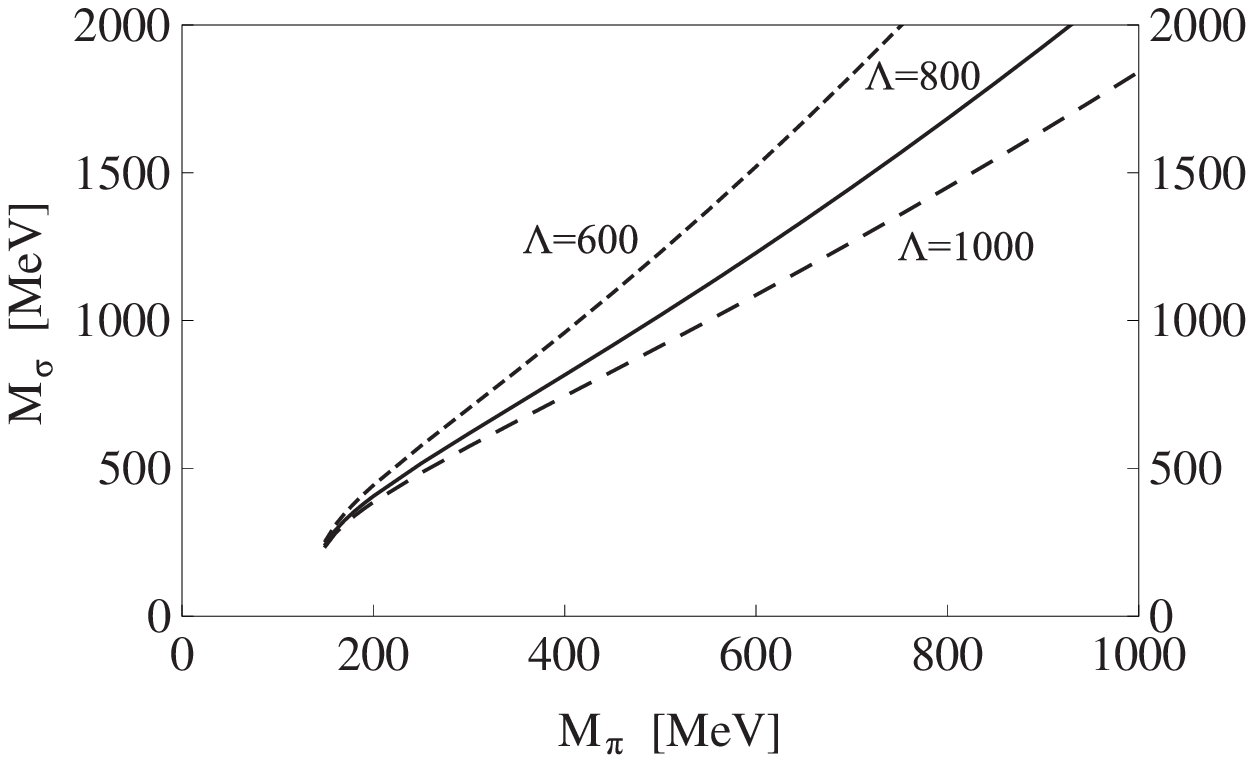}
\figcaption{\label{fig:gapESB} The relation of the dressed $\sigma$ mass $M_\sigma$ and the dressed $\pi$ mass $M_\pi$ for the case of explicit chiral symmetry breaking. The short-dashed, solid and long-dashed curves are for $\Lambda = 600, 800$ and $1000$\,MeV, respectively.}
\end{center}

The two cases provide essentially the same results except for the existence of some threshold effects. We show $\mu_0$ and the dressed masses of $\sigma$ and $\pi$ as functions of $\lambda_0$ in Fig.\,\ref{fig:ESB} for $\Lambda = 800$\,MeV.  These results are similar to those obtained in the chiral limit, except that the dressed pion mass has a threshold value due to the explicit chiral symmetry breaking term.

\end{multicols}
\ruleup
\begin{center}
\includegraphics[width=8cm]{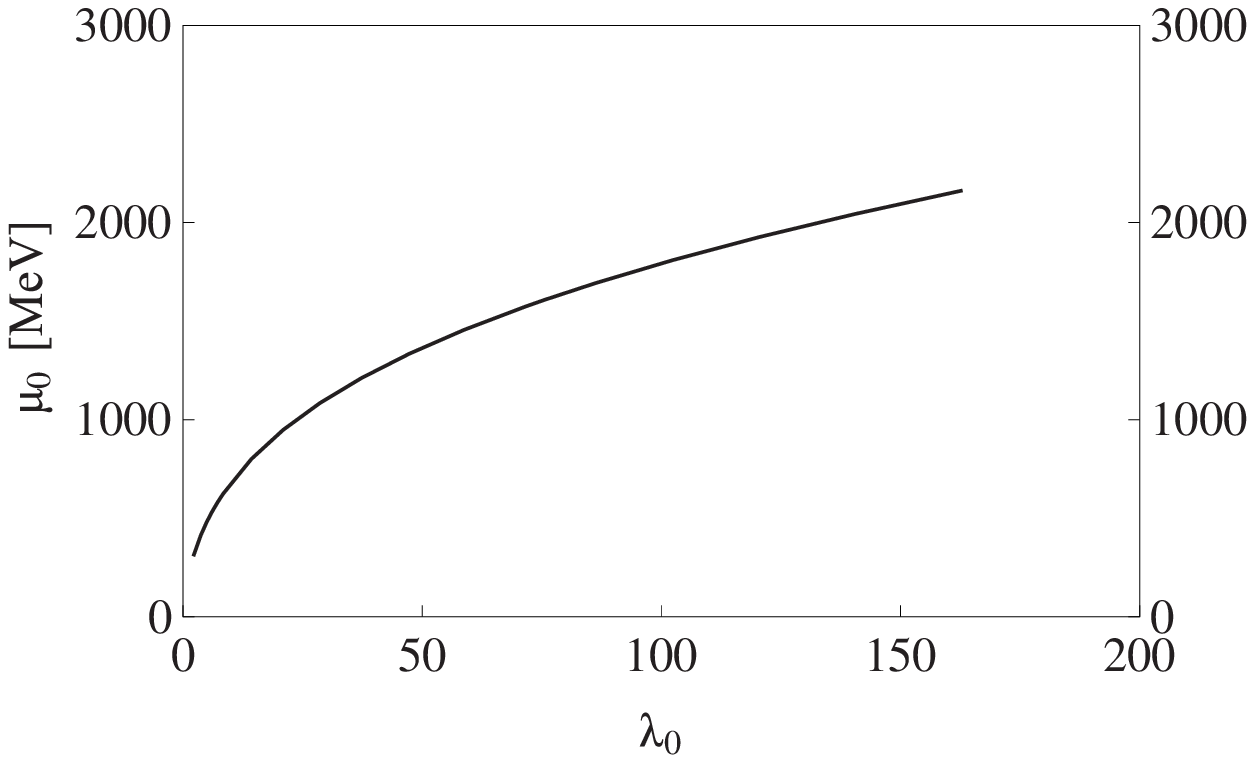}
\includegraphics[width=8cm]{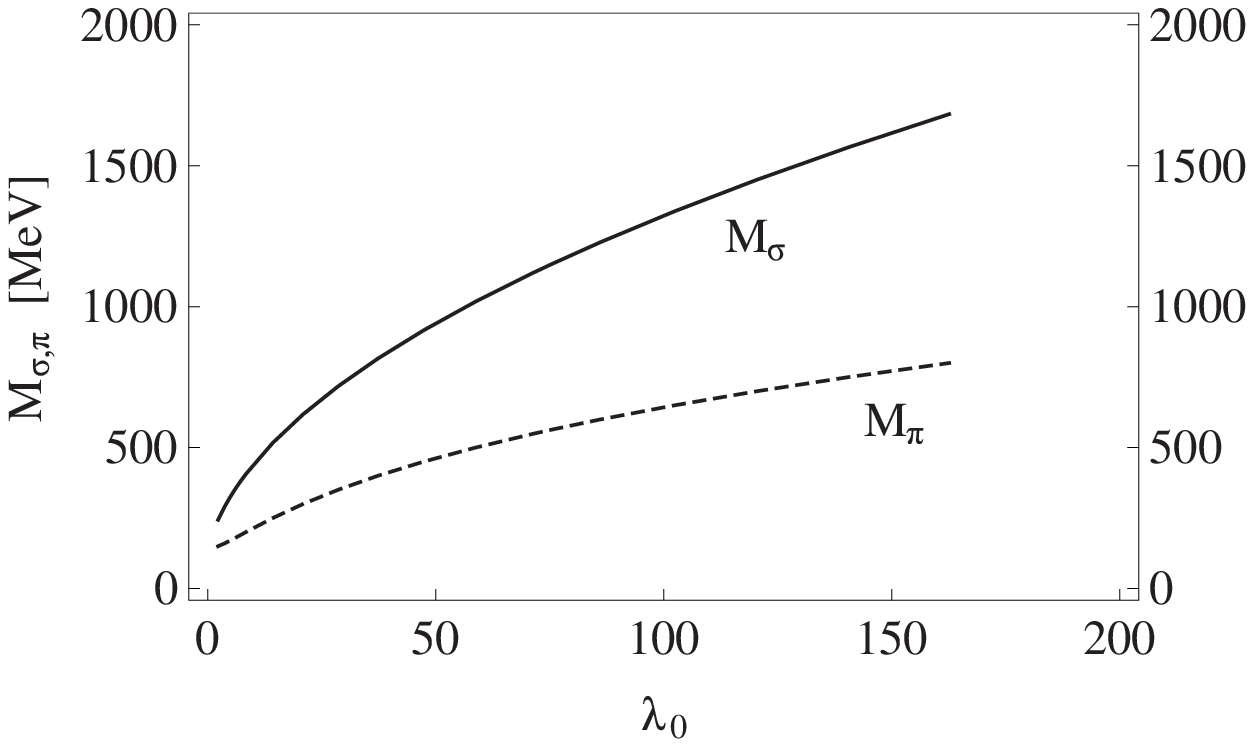}
\figcaption{\label{fig:ESB} $\mu_0$ and the dressed $\sigma$ and $\pi$ masses as functions of $\lambda_0$ for $\Lambda = 800$\,MeV. The calculations are done with an explicit chiral symmetry breaking parameter $\varepsilon \neq 0$.}
\end{center}
\ruledown

\begin{multicols}{2}

Using the Bethe-Salpeter equations \eqref{eq:BSpion} and \eqref{eq:BSsigma}, we can obtain the physical pion and sigma masses, as shown in Fig.\,\ref{fig:BSESB} as functions of the dressed pion mass $M_\pi$. The physical pion mass $m_{\pi}$ is finite and close to its physical value 138\,MeV. The physical sigma mass $m_{\sigma}$ increases with $M_{\pi}$.

\begin{center}
\includegraphics[width=8cm]{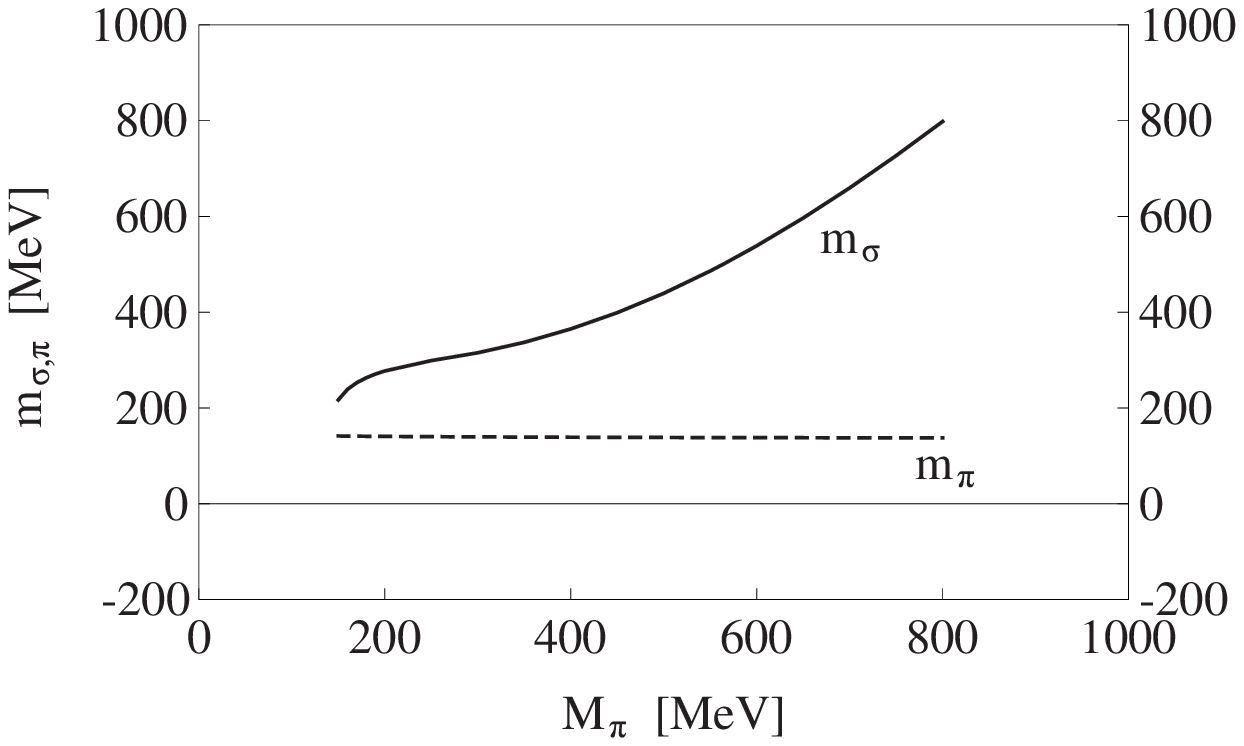}
\figcaption{\label{fig:BSESB} Physical $\pi$ and $\sigma$ masses, $m_{\sigma}$ and $m_{\pi}$, as functions of $M_\pi$ obtained by solving the Bethe-Salpeter equations (\ref{eq:BSpion}) and (\ref{eq:BSsigma}). The calculations are done with an explicit chiral symmetry breaking parameter $\varepsilon \neq 0$.}
\end{center}

We choose the following parameters in the linear $\sigma$ model Lagrangian for the explicit chiral symmetry breaking case:
\begin{eqnarray}
\nonumber \varepsilon &=& 142^2 \times 93.0~ {\rm MeV}^3=1.86 \times10^{6}~{\rm MeV}^{3}  \, ,
\\ \label{eq:EBparameter}
\lambda_0 &=& 75.5 \, ,
\nn
\mu_0 &=& 1610 \,{\rm MeV} \, ,
\nn
\Lambda &=& 800 \,{\rm MeV} \, .
\end{eqnarray}
These values provide
\begin{eqnarray}
v(T=0) &=& f_\pi = 93.0 \,{\rm MeV} \, ,
\nn
M_\sigma(T=0) &=& 1150 \,{\rm MeV} \, ,
\nn
M_\pi(T=0) &=& 564 \,{\rm MeV} \, ,
\\ \label{eq:EBvalues}
m_\sigma(T=0) &=& 500 \,{\rm MeV} \, ,
\nn
\nonumber m_\pi(T=0) &=& 138 \,{\rm MeV} \, .
\end{eqnarray}

\section{Finite Temperature Analyses}
\label{sec:finite}

In this section we study the behavior of hadron properties at finite temperature. We shall use the parameters fixed in the previous section, which are obtained at zero temperature, and study both the chiral limit ($\varepsilon=0$) and the explicit chiral symmetry breaking case with $\varepsilon \neq 0$.  By doing this we shall see how the chiral symmetry is recovered at finite temperature.


\subsection{Chiral limit ($\varepsilon=0$)}

We discuss first the case of chiral limit ($\varepsilon=0$). The sigma mean field value $v$ is shown in Fig.\,\ref{fig:fvchi} as a function of temperature $T$. It starts at $v= 93$\,MeV, and stays around this value for a while and suddenly drops to zero around the critical temperature $T_C = 194$\,MeV.  At this temperature, the free energies at $v=0$ and $v\sim 80$\,MeV agree with each other and the corresponding free energy is ${\cal F}\sim -5\times 10^{8}$\,MeV$^4$.  There are multiple minima near the critical temperature indicating first order phase transition.
This is caused by the use of large coupling constant, $\lambda_0=83.6$, in order to reproduce the sigma meson mass at zero temperature.

\begin{center}
\includegraphics[width=8cm]{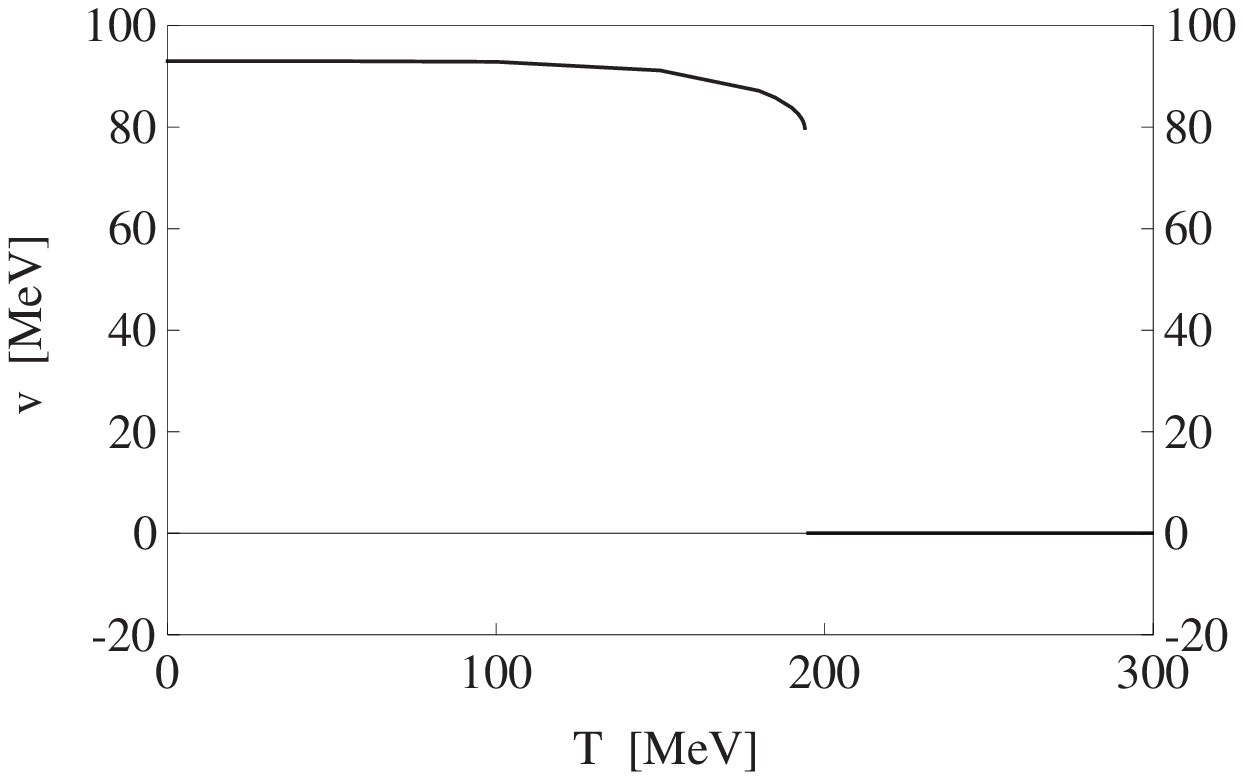}
\figcaption{\label{fig:fvchi} The sigma mean field value $v$ as a function of the temperature $T$ in the chiral limit.  The clear discontinuity of $v$ around $T=194$\,MeV indicates a first order phase transition.}
\end{center}

The dressed sigma and pion masses, $M_\sigma(T)$ and $M_\pi(T)$, are shown in Fig.\,\ref{fig:mmasschi} also as functions of $T$.  $M_{\sigma}$ stays at a large value around 1200~MeV,  and starts to drop around $T_C$, while it increases above this temperature. The dressed pion mass $M_{\pi}$ is almost constant around 600~MeV until $T_C$, and also increases above this temperature.  We note that above the critical temperature $T_C = 194$\,MeV the chiral symmetry is recovered, and the dressed sigma and pion masses $M_{\sigma}$ and $M_{\pi}$ coincide in the chiral limit.

\begin{center}
\includegraphics[width=8cm]{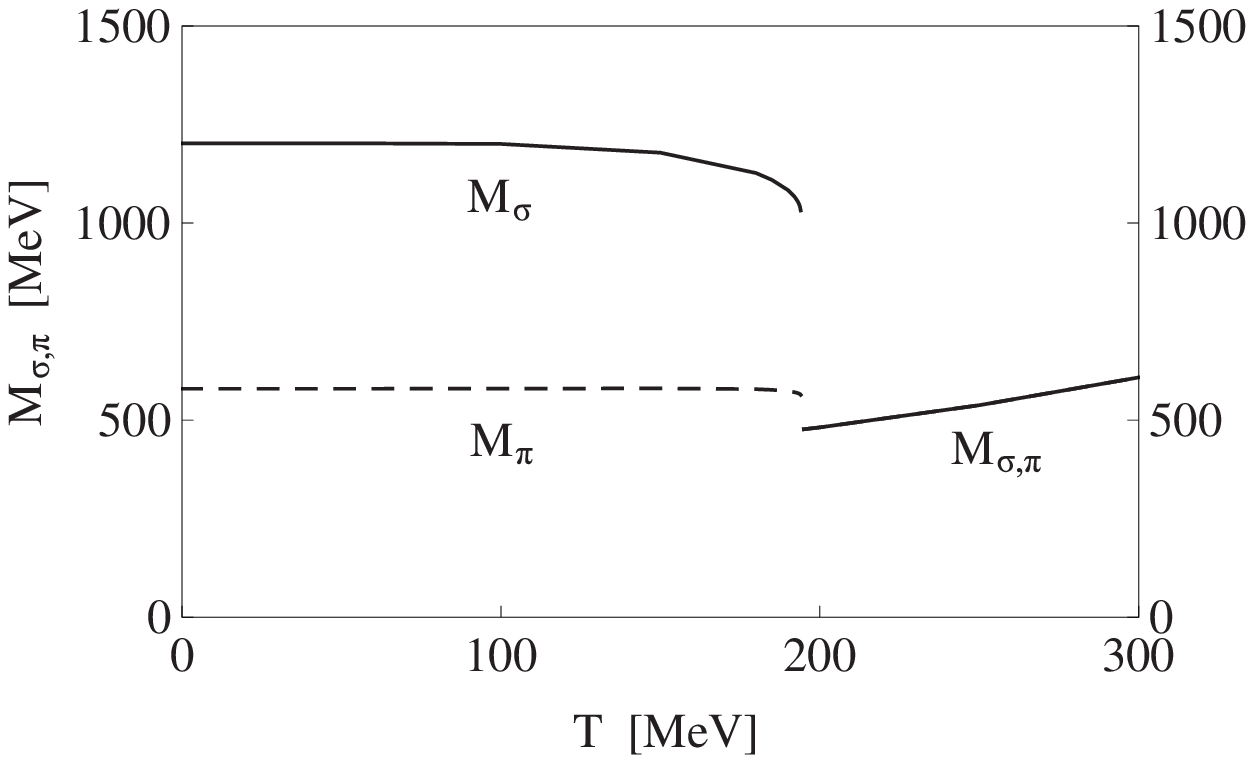}
\figcaption{\label{fig:mmasschi} The dressed sigma and pion masses, $M_{\sigma}$ and $M_{\pi}$, as functions of the temperature $T$. $M_{\sigma}$ and $M_{\pi}$  coincide above the critical temperature $T_C= 194$\,MeV, where the chiral symmetry is recovered $v=0$. The calculations are done in the chiral limit.}
\end{center}

We solve the Bethe-Salpeter equations\, \eqref{eq:BSpion} and \eqref{eq:BSsigma} to calculate the physical sigma and pion masses. The resulting masses are plotted in Fig.\,\ref{fig:physmasschi} as functions of $T$. We see that the physical pion mass stays at zero until $T_C = $194\,MeV, which means that the Nambu-Goldstone theorem is recovered as long as the chiral symmetry is spontaneously broken ($v=$ finite). When the chiral symmetry is recovered above $T_C$, the physical pion mass is not zero any more. It coincides with the physical sigma mass, and they both increase with the temperature $T$.  These physical masses correspond to the dressed masses in the chiral symmetric phase.

\begin{center}
\includegraphics[width=8cm]{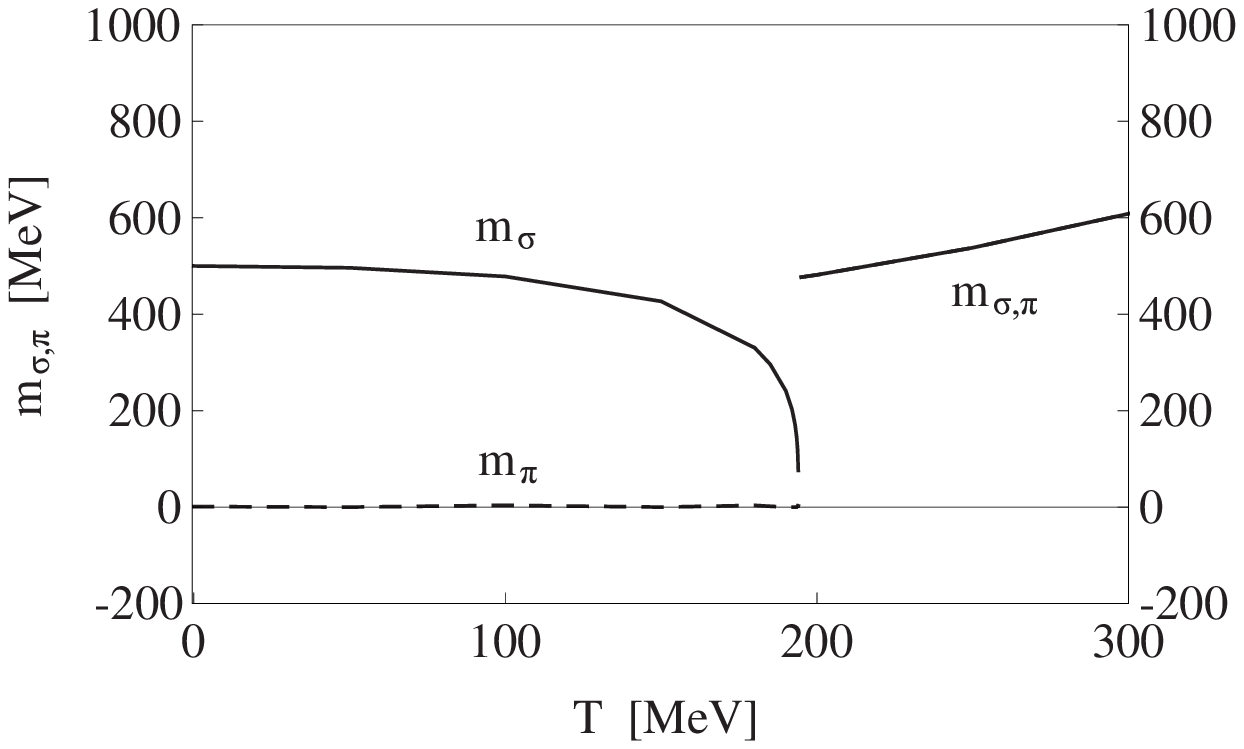}
\figcaption{\label{fig:physmasschi} The physical sigma and pion masses, $m_{\sigma}$ and $m_{\pi}$, as functions of the temperature $T$.  The pion mass $m_{\pi}$ stays at zero mass until the critical temperature.  The sigma mass $m_{\sigma}$  is finite and drops to a small value at the critical temperature.  They coincide above the critical temperature $T_C= 194$\,MeV, where the chiral symmetry is recovered. The calculations are done in the chiral limit.}
\end{center}

In order to understand the behavior of the physical mass, we multiply $m_{\pi}^2-M_{\pi}^2$ to the pole condition equation $1-V_{\sigma\pi\rightarrow \sigma\pi}(s)G_{\sigma\pi\rightarrow \sigma\pi}(s)=0$.  We then find the following equation:
\beq
m_{\pi}^2=M_{\pi}^2+\frac{4\lambda_0^2v^2G_{\sigma\pi\rightarrow\sigma\pi}(m_{\pi}^2)}{1-2\lambda_0G_{\sigma\pi\rightarrow\sigma\pi}(m_{\pi}^2)}~.
\eeq
This equation indicates that in the range $m_{\pi} < M_{\pi}$, $m_{\pi}$ goes to $M_{\pi}$ as $v\rightarrow 0$.  This means that the BS equation does not provide a bound state of pion-sigma composite above the critical temperature.  Hence, the pion mass becomes the thermal mass, which increases with the temperature.

We make a comment on the cut-off momentum dependence of the calculated results.  We have been using the cut-off momentum of $\Lambda=800$\,MeV.  In order to see the cut-off momentum dependence, we calculated the case of $\Lambda=1000$\,MeV.  Definitely, all the numbers changed from the original case.  We used a constraint on the parameters that the sigma physical mass was $m_\sigma=500$\,MeV.  We found that the behavior of phase transition was almost identical to the original case where the sigma mass $m_\sigma$ and the sigma mean field value $v$ dropped with temperature and the critical temperature was essentially unchanged from the original case of $\Lambda=800$\,MeV.

\subsection{Explicit chiral symmetry breaking case ($\varepsilon \ne 0$)}

We repeat the same calculation for the case of explicit chiral symmetry breaking.
We use the mean field equation and mass-gap equations, Eqs.\,\eqref{eq:veva}, \eqref{eq:gapa} and \eqref{eq:gapb} to calculate the sigma mean field value $v$ and the dressed sigma and pion masses, $M_{\sigma}$ and $M_{\pi}$. The sigma mean field value $v$ is shown in Fig.\,\ref{fig:mean1} as a function of the temperature $T$. It gradually decreases as the temperature increases until the critical temperature $T_C=198$\,MeV, and suddenly drops to a small finite value. After this critical temperature, it gradually decreases towards zero due to explicit chiral symmetry breaking. The dressed sigma and pion masses are shown in Fig.\,\ref{fig:nmass} as functions of $T$. Their behaviors are similar to those in the chiral limit. Again these behaviors indicate a first order phase transition even for the explicit chiral symmetry breaking case opposing to the MF case.

\begin{center}
\includegraphics[width=8cm]{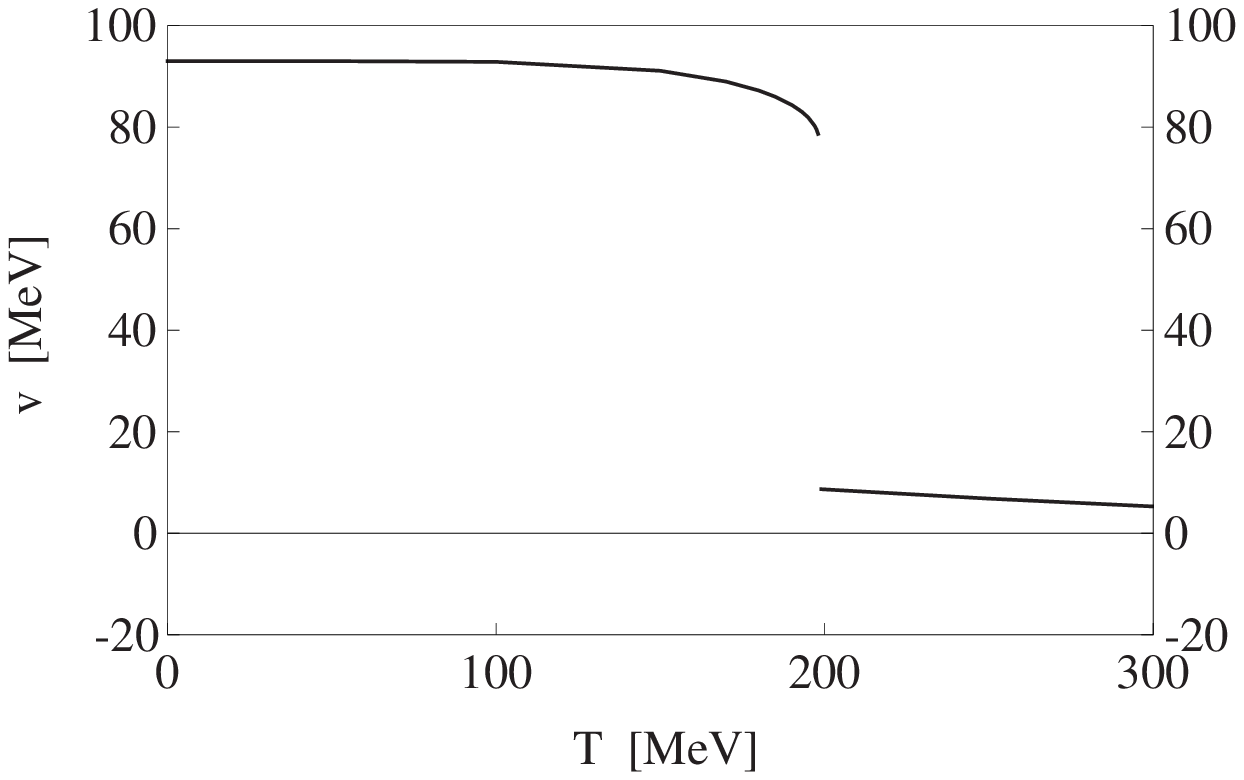}
\figcaption{\label{fig:mean1} The sigma mean field value $v$ as a function of the temperature $T$. The calculations are done with an explicit chiral symmetry breaking parameter $\varepsilon \neq 0$.}
\end{center}

\begin{center}
\includegraphics[width=8cm]{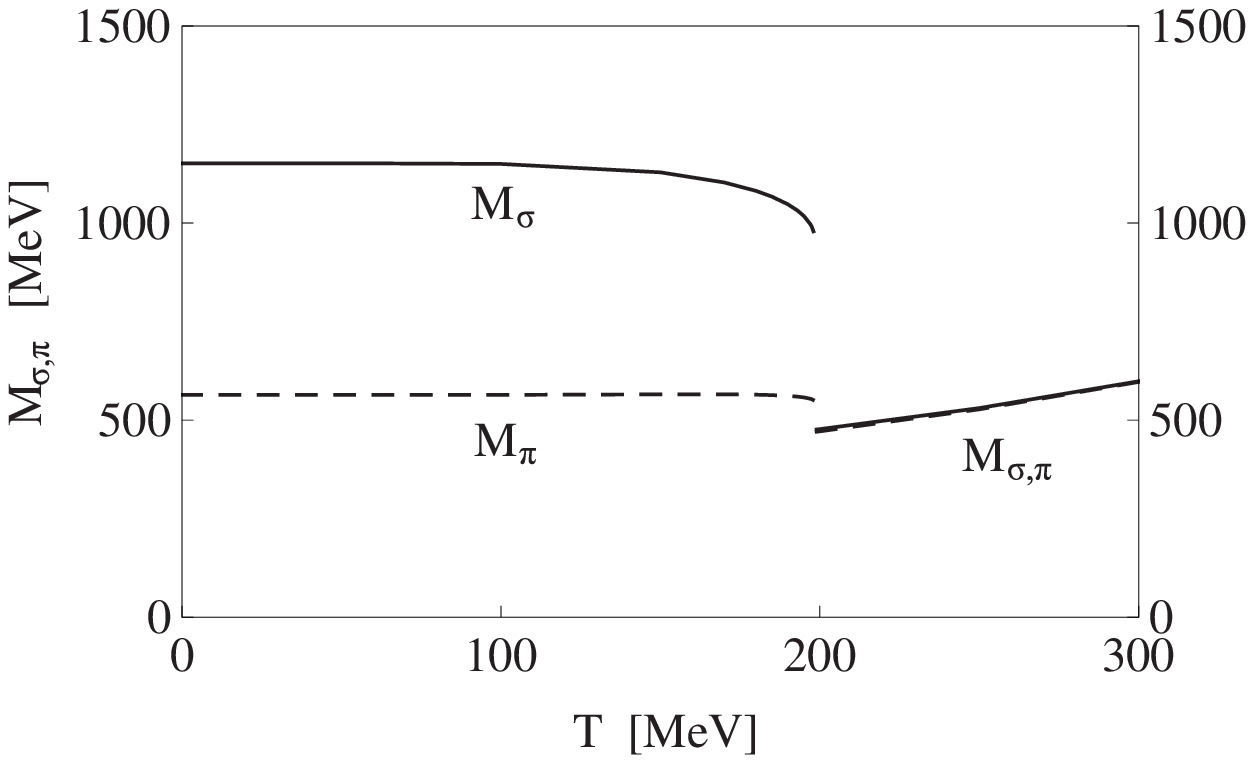}
\figcaption{\label{fig:nmass} The dressed sigma and pion masses, $M_{\sigma}$ and $M_{\pi}$, as functions of the temperature $T$. They almost coincide above the critical temperature $T_C= 198$\,MeV, where the chiral symmetry is almost recovered. The calculations are done with an explicit chiral symmetry breaking term $\varepsilon \neq 0$.}
\end{center}

We solve then the Bethe-Salpeter equations \eqref{eq:BSpion} and \eqref{eq:BSsigma} to calculate the physical sigma and pion masses, $m_{\sigma}$ and $m_{\pi}$. The results are shown in Fig.\,\ref{fig:physmass}. In this case, the pion mass stays around $m_{\pi}=138$\,MeV below $T_C=198$\,MeV and then increases suddenly to a large value. Again the physical sigma and pion masses have almost the same values above $T_C$, where the chiral symmetry is almost recovered.  The reason of this behavior is similar to the case of the chiral limit.  The meson masses above the critical temperature correspond to the thermal masses.

\begin{center}
\includegraphics[width=8cm]{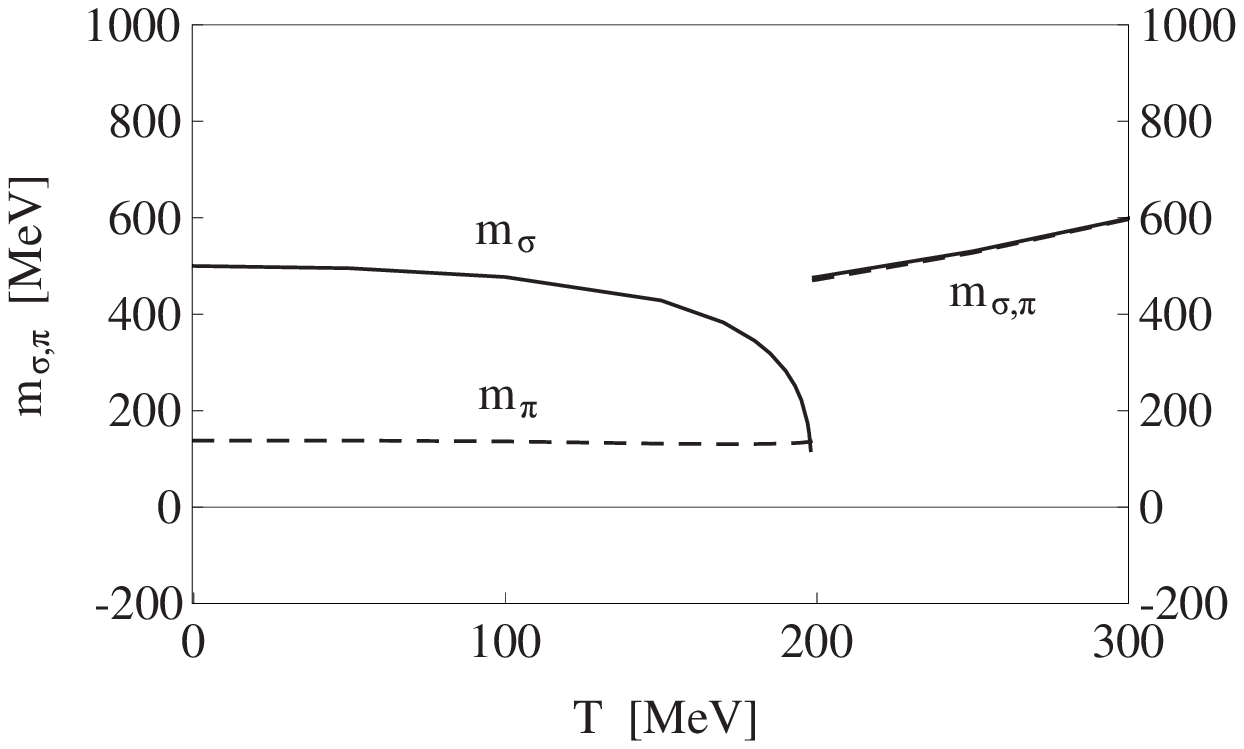}
\figcaption{\label{fig:physmass} The physical sigma and pion masses, $m_{\sigma}$ and $m_{\pi}$, as functions of the temperature $T$. They almost coincide above the critical temperature $T_C= 198$\,MeV, where the chiral symmetry is recovered. The calculations are done with an explicit chiral symmetry breaking parameter $\varepsilon \neq 0$.}
\end{center}

We compare our results with those of Tsue and Matsuda~\cite{tsue08}.  In their calculation, they dropped the contribution of $\sigma-\sigma$ excitations for the physical $\sigma$ mass in the BS equation, by saying that the threshold energy of $\sigma-\sigma$ excitations is much larger than the value of the cut-off momentum.  This made the physical $\sigma$ state to appear at higher energy by about 400\,MeV as compared to our case, where the $\sigma-\sigma$ excitations were included explicitly in the BS equation.  Hence, they could take smaller coupling constants with smaller cut-off momentum in order to obtain the physical $\sigma$ mass at $m_\sigma=500$\,MeV.  This procedure made the loop contributions much smaller than our case.  In this case, they found a second order phase transition for the explicit chiral symmetry breaking case as the MF approximation.  In our case, we prefer to include explicitly the $\sigma-\sigma$ excitations in the BS equation.  In construction of the linear $\sigma$ model, the higher energy mode has been integrated out at the quark-gluon level in QCD and the resulting interactions among mesons after bosonization by integrating out quark fields and the low energy gluon fields take care of the quark-gluon dynamics~\cite{kondo11}.  This procedure naturally introduces a certain momentum scale due to the compositeness of mesons.  Hence, we take all the excitation modes produced in the linear $\sigma$ model up to a certain momentum in terms of the cut-off momentum $\Lambda$.  In addition, the Goldstone theorem to be recovered needs the excitations of $\sigma-\pi$ states, which have similar excitation energies as the $\sigma-\sigma$ states.  Instead, we expect some states associated with the $\chi$ field due to the pure gluon dynamics around 1$\sim$ 2\,GeV as intruder states in the meson spectrum.  As for the order of phase transition, even for the explicit symmetry breaking case the contribution of the explicit symmetry breaking term to the free energy near the critical point is about one order smaller than other terms and it remains a first order phase transition.  Hence, the order of the phase transition was not influenced by the explicitly symmetry breaking term.  If we increase the magnitude of $\varepsilon$, the phase transition has a tendency to become second order as the gap of the order parameter $v$ at the critical temperature becomes smaller with $\varepsilon$.

\section{Summary and Conclusion}
\label{sec:summary}

We have studied the hadron properties using the O(4) linear $\sigma$ model, where we have used the Gaussian functional (GF) method to improve the mean field approximation by including fluctuations around the mean field values. The GF method provides a finite pion mass for the case of spontaneous chiral symmetry breaking in the chiral limit, and does not respect the Nambu-Goldstone theorem at this stage. To this end, we have solved the Bethe-Salpeter (BS) equations for the physical sigma and pion masses.  We have found that the physical pion mass drops to zero by solving the BS equation after the GF method, indicating that the sigma model respects the Nambu-Goldstone theorem at this stage~\cite{nakamura01,chen10}.

We chose the parameters of the sigma model so as to get $m_{\sigma}=500$\,MeV. In the case of chiral limit, the physical pion mass is always zero due to the Nambu-Goldstone theorem; while in the case of explicit chiral symmetry breaking, it is not zero any more. Accordingly in the latter case there is an extra parameter to represent explicit chiral symmetry breaking, $\varepsilon \neq 0$.  It is fine-tuned so that the physical pion mass becomes $m_{\pi}=138$\,MeV. These parameters are shown in Eqs.~\eqref{eq:NBparameter} and \eqref{eq:EBparameter}.  We found that the change from the dressed sigma and pion masses to their physical masses is very large and of order of 600$\sim$700\,MeV. Therefore, in order to obtain a reasonable physical sigma mass around 500\,MeV, we ought to take a large dressed sigma mass around 1200\,MeV, and consequently a large coupling constant around $\lambda_{0}\approx80$.  This large coupling of mesons causes large changes of sigma and pion masses from their dressed values to physical values.  The sigma meson is considered as a strongly correlated meson-meson state, that indicates it as a four quark state~\cite{chen10a}.
Moreover, this coupling constant is so large that the order of the phase transition at finite temperature is of first order.

Using these parameters, i.e. Eqs.~\eqref{eq:NBparameter} and \eqref{eq:EBparameter}, which are fixed at zero temperature, we have calculated the sigma mean field value and the sigma and pion masses as functions of temperature by using the GF + BS method, for the chiral limit and explicit chiral symmetry breaking case. For the latter case the pion mass stays at around $m_{\pi}=138$\,MeV until the critical temperature $T_C=198$\,MeV. At this temperature, the order parameter, i.e. the sigma mean field value (condensate) $v$, changes rapidly from $v \approx 93$\,MeV to a small value, and then gradually decreases to zero with temperature. Reflecting this behavior, both the dressed and the physical pion masses jump up rapidly at $T_C=198$\,MeV, to large values, corresponding to the sigma masses. This behavior is caused by the large coupling constant $\lambda_{0}$ for the meson interactions, when we consider fluctuations around the mean field values using the GF method, and the physical pion and sigma mesons gain large binding energies of dressed pion and sigma meson obtained by solving the BS equations in the case of chiral symmetry breaking below the critical temperature.  On the other hand, above the critical temperature the BS equations do not provide bound states in both channels and the sigma and pion masses become those of the thermal masses.

In general discussion, the linear $\sigma$ model with O(4) symmetry provided second-order phase transition at large temperature.  However, when the critical temperature is close to the hadron scale, we have to introduce a model assumption.  In the discussion by Ogure and Sato, they made one assumption for the self energy of the Goldstone boson to have zero mass in the spontaneously broken phase~\cite{ogure99}.  On the other hand, probably due to the large coupling constant we are using, the present GF+BS formalism provided a first-order phase transition, where we have considered the linear $\sigma$ model as a low energy effective theory of QCD, and introduced the cut-off momentum at some physical scale.  There is further a discussion on the setting-sun diagram, which may become important around the phase transition temperature \cite{pilaftsis13}.  It is our future work to make clear this point.  It is also very interesting to study the properties of mesons for the SU(3) flavor case using the GF+BS method.

\acknowledgments{We thank Prof. V.~Dmitra\u{s}inovi\'{c} for fruitful discussions and comments and S.~I. acknowledges Dr. H. Iida for useful comments.
This work is supported by the JSPS research grant (S): 21540267.  H.X.C and L.S.G are supported by the Fundamental Research Funds for the Central Universities.  S.I is supported by the Grant for Scientific Research [Priority Areas ``New Hadrons'' (E01:21105006), (C) No.\,23540306] from the Ministry of Education, Culture,Science and Technology (MEXT) of Japan.}

\end{multicols}

\vspace{15mm}

\begin{multicols}{2}

\end{multicols}

\vspace{-1mm}
\centerline{\rule{80mm}{0.1pt}}
\vspace{2mm}

\begin{multicols}{2}

\end{multicols}

\clearpage

\end{document}